\documentclass[twocolumn,tighten]{aastex62}

\usepackage{amsmath}	
\usepackage{amssymb}	
\usepackage[figuresright]{rotating}

\usepackage{soul, xcolor}

\newcommand{\ron}{\color{black}}   

\setstcolor{magenta}

\newcommand{\NSecTotalRRMdwarf}{371}

\newcommand{\NSecTotalFastMdwarf}{17}

\usepackage{underscore}

\newcommand{\guentherinprep}{(G{\"u}nther \& Daylan, in prep.)}

\newcommand{\corr}[1]{\textcolor{magenta}{#1}}

\received{Jan xx th, 2019}
\revised{---}
\accepted{---}
\submitjournal{ApJ}

\shorttitle{Complex Rotation Profiles}
\shortauthors{Zhan et al.}

\begin{document}

\title{Complex Rotational Modulation of Rapidly Rotating M-Stars Observed with TESS}

\author[0000-0002-4142-1800]{Z.~Zhan}
\affiliation{Department of Earth, Atmospheric, and Planetary Sciences, M.I.T., Cambridge, MA 02139, USA}
\correspondingauthor{Z. ~Zhan}
\email{zzhan@mit.edu}

\author[0000-0002-3164-9086]{M.~N.~G{\"u}nther}
\affiliation{Department of Physics and Kavli Institute for Astrophysics and Space Research, M.I.T., Cambridge, MA 02139, USA}
\affiliation{Juan Carlos Torres Fellow}

\author[0000-0003-3182-5569]{S.~Rappaport}
\affiliation{Department of Physics and Kavli Institute for Astrophysics and Space Research, M.I.T., Cambridge, MA 02139, USA}

\author[0000-0003-3669-7201]{K.~Ol\'ah}
\affiliation{Konkoly Observatory, Research Centre for Astronomy and Earth Sciences, HAS, H-1121 Budapest, Konkoly Thege M. u. 15-17, Hungary}

\author[0000-0003-3654-1602]{A.~Mann}
\affiliation{Department of Physics and Astronomy, University of North Carolina at Chapel Hill, Chapel Hill, NC 27599, USA}

\author[0000-0001-8172-0453]{A.~M.~Levine}
\affiliation{Department of Physics and Kavli Institute for Astrophysics and Space Research, M.I.T., Cambridge, MA 02139, USA}

\author[0000-0002-4265-047X]{J.~Winn}
\affiliation{Department of Astrophysical Sciences, Princeton University, Princeton, NJ 08544, USA}

\author[0000-0002-8958-0683]{F.~Dai}
\affiliation{Department of Astrophysical Sciences, Princeton University, Princeton, NJ 08544, USA}
\affiliation{Department of Physics and Kavli Institute for Astrophysics and Space Research, M.I.T., Cambridge, MA 02139, USA}

\author[0000-0002-4891-3517]{G.~Zhou}
\affiliation{Center for Astrophysics $|$ Harvard $\&$ Smithsonian, 60 Garden St., Cambridge, MA 02138, USA}
\affiliation{Hubble Fellow}

\author[0000-0003-0918-7484]{Chelsea X.~Huang}
\affiliation{Department of Physics and Kavli Institute for Astrophysics and Space Research, M.I.T., Cambridge, MA 02139, USA}
\affiliation{Juan Carlos Torres Fellow}

\author[0000-0002-0514-5538]{L.G.~Bouma}
\affiliation{Department of Astrophysical Sciences, Princeton University, Princeton, NJ 08544, USA}

\author[0000-0002-6194-043X]{M.J.~Ireland}
\affiliation{Research School of Astronomy and Astrophysics, Australian National University, Canberra, ACT 2611, Australia}

\author[0000-0003-2058-6662]{G.~Ricker}
\affiliation{Department of Physics and Kavli Institute for Astrophysics and Space Research, M.I.T., Cambridge, MA 02139, USA}

\author[0000-0001-6763-6562]{R.~Vanderspek}
\affiliation{Department of Physics and Kavli Institute for Astrophysics and Space Research, M.I.T., Cambridge, MA 02139, USA}

\author[0000-0001-9911-7388]{D.~Latham}
\affiliation{Center for Astrophysics $|$ Harvard $\&$ Smithsonian, 60 Garden St., Cambridge, MA 02138, USA}

\author[0000-0002-6892-6948]{S.~Seager}
\affiliation{Department of Earth, Atmospheric, and Planetary Sciences, M.I.T., Cambridge, MA 02139, USA}
\affiliation{Department of Physics and Kavli Institute for Astrophysics and Space Research, M.I.T., Cambridge, MA 02139, USA}
\affiliation{Department of Aeronautical and Astronautical Engineering, M.I.T., Cambridge, MA 02139, USA}

\author[0000-0002-4715-9460]{J.~Jenkins}
\affiliation{NASA Ames Research Center, Moffett Field, CA 94035, USA}

\author[0000-0003-1963-9616]{D.~A.~Caldwell}
\affiliation{SETI Institute, Mountain View, CA  94043, USA}

\author{J.~P.~Doty}
\affiliation{Noqsi Aerospace Ltd, 15 Blanchard Avenue, Billerica, MA 01821, USA}

\author[0000-0002-2482-0180]{Z.~Essack}
\affiliation{Department of Earth, Atmospheric, and Planetary Sciences, M.I.T., Cambridge, MA 02139, USA}

\author{G.~Furesz}
\affiliation{Department of Physics and Kavli Institute for Astrophysics and Space Research, M.I.T., Cambridge, MA 02139, USA}

\author[0000-0003-4724-745X]{M.~E.~R.~Leidos}
\affiliation{NASA Ames Research Center, Moffett Field, CA 94035, USA}

\author[0000-0002-4829-7101]{P.~Rowden}
\affiliation{School of Physical Sciences, The Open University, Milton Keynes MK7 6AA, UK}

\author[0000-0002-6148-7903]{J.~C.~Smith}
\affiliation{NASA Ames Research Center, Moffett Field, CA 94035, USA}
\affiliation{SETI Institute, Mountain View, CA  94043, USA}

\author[0000-0002-3481-9052]{K.~G.~Stassun}
\affiliation{Department of Physics $\&$ Astronomy, Vanderbilt University, 6301 Stevenson Center Ln., Nashville, TN 37235, USA}

\author{M.~Vezie}
\affiliation{Department of Physics and Kavli Institute for Astrophysics and Space Research, M.I.T., Cambridge, MA 02139, USA}




\begin{abstract}
We have searched for short periodicities in the light curves of stars with $T_{\rm eff}$ cooler than 4000 K made from 2-minute cadence data obtained in {\em TESS} sectors 1 and 2.   Herein we report the discovery of 10 rapidly rotating M-dwarfs with highly structured rotational modulation patterns among {\NSecTotalRRMdwarf} M dwarfs found to have rotation periods less than 1 day.  Star-spot models cannot explain the highly structured periodic variations which typically exhibit between 10 and 40 Fourier harmonics.  A similar set of objects was previously reported following {\em K2} observations of the Upper Scorpius association (Stauffer et al. 2017).  We examine the possibility that the unusual structured light-curves could stem from absorption by charged dust particles that are trapped in or near the stellar magnetosphere.  We also briefly explore the possibilities that the sharp structured features in the lightcurves are produced by extinction by coronal gas, by beaming of the radiation emitted from the stellar surface, or by occultations of spots by a dusty ring that surrounds the star.  The latter is perhaps the most promising of these scenarios. Most of the structured rotators display flaring activity, and we investigate changes in the modulation pattern following the largest flares. As part of this study, we also report the discovery of {\NSecTotalFastMdwarf} rapidly rotating M-dwarfs with rotational periods below 4 hr, of which the shortest period is 1.63 hr.
\end{abstract}





\section{Introduction}
\label{sec:intro}

Young M stars often are found to rotate rapidly, i.e., with periods $\lesssim 2$ days, and then are seen to be `active' in a number of different respects.  
The spectra of such stars may contain hydrogen as well as Ca H\&K emission lines \citep{pettersen89}; flares on such stars are common (e.g. \citealt{davenport19,mondrik19,gunther19}); and the stellar surfaces often have spots that result in large-amplitude rotational modulations (e.g., EY Dra, \citealt{vida10}; GJ 65A and 65B, \citealt{barnes17}).  Even though the spots may be distributed around the stellar surface, the rotational modulation patterns are usually fairly simple, i.e., they may be adequately represented by just a few Fourier components.  A small percentage of rotationally modulated stars, e.g. KIC 7740983 \citep{rappaport14} and HSS 348 \citep{strassmeier17}, exhibit more complex rotational modulations; even these can often be ascribed to particular sizes and arrangements of spots.  Sometimes two distinct rotation frequency signatures are seen, indicating the presence of two `twin-like' M stars \citep{rappaport14}. 

Two other quite distinct categories of young rotating M stars have recently been identified.  First, there is a class of young late-type stars that exhibit `dipping activity' (see, e.g., \citealt{cody14}; \citealt{ansdell16}). Such stars can exhibit dips with depths that range from $\sim$5\% to 50\%, and that may recur periodically, quasi-periodically, or without any apparent underlying pattern. These stars typically exhibit strong emission in the WISE 3 and 4 bands, indicating that dusty disks are present. Second, there is a class of stars found by \citet{stauffer17,stauffer18} that may be related to the dippers, that the authors describe as ``rapidly rotating weak-line T Tau stars with orbiting clouds of material at the Keplerian co-rotation radius''.  These stars were found in the Upper Scorpius and $\rho$ Oph star forming associations, and exhibit highly structured rotational modulations that, according to the authors, cannot be explained by star spots.  

In this work we present the discovery with the Transiting Exoplanet Survey Satellite {\em TESS} \citep{ricker15} of 10 rapidly rotating M stars in the Southern Hemisphere with quite complex rotational modulation patterns (the `10 complex rotators').  Half of these stars are in the star-forming stellar association Tucana Horologium (`THA').  In Section~\ref{sec:observations} we discuss the data from Sectors 1 and 2 that we searched for interesting rapid rotators.  In Section~\ref{sec:methods} we describe the Fourier transform survey of all the 2-minute cadence targets in Sectors 1 and 2, and discuss the Fourier amplitude spectra, the rotational modulation patterns, and other characteristics of the 10 complex rotators.  Low and medium resolution spectra of a sample of the stars are presented in Section~\ref{sec:spectra}.   We attempt to model the light curves with a starspot modeling code in Section~\ref{sec:spots}, but conclude that such a model falls short of explaining the observations. In Section~\ref{sec:magnetic} we discuss another model involving charged dust particles trapped in the stellar magnetosphere, but find that this scenario also has difficulties.  A model, perhaps the most promising one considered here, that involves a spotted host star rotating obliquely inside a dusty ring is presented in Section~\ref{sec:ring}.
In Section~\ref{sec:beaming} we raise the possibility that the highly structured rotational modulations might be due to beamed emission. Changes in the structured modulation pattern around the time of a superflare in one of the rapid rotators are discussed in Section~\ref{sec:flare}. In Section~\ref{sec:associations} we discuss the stellar associations where these 10 stars are found, and their likely ages.  We summarize our work and discuss some related issues in Section~\ref{sec:summary}.

\begin{table*}
\centering
\null{} \hspace{2cm}
\caption{M Stars With Rotation Periods $P < 4$ hr in {\em TESS} Sectors 1 \& 2}
\begin{tabular}{lccccccc}
\hline
\hline
 TIC & RA J2000 &Dec J2000 &   Period$^a$ (hr)  & $T_{\rm eff}$$^b$ K & $\Omega/\Omega_{\rm crit}$$^c$  & $M_{\rm K}$$^d$ & {\ron Sectors$^e$} \\
\hline
204294226 	& 345.3914  	& $-22.0765$ & $1.631(2)$ & 3383  	& 0.572 & 6.864 & 2\\ 
332507607 	& 350.5944 	& $-26.7901$ & $2.100(1)$ & 2877	& 0.332 & 7.719 & 2\\  
261089844   	& ~\,81.5581 	& $-83.6062$ & $2.151(1)$ & 3705 	& ...       & ...       & 1\\ 
231838187   	& ~\,24.9002  	& $-$64.9294 & $2.306(1)$ & 2830 	& 0.275 & 7.99   & 1, 2\\  
120522722   	& ~\,~\,2.8139 	& $-$37.9487 & $2.409(1)$ & 2771  	& 0.273 & 7.891 & 2\\ 
~\,12419331 	& ~\,~\,2.0585 	& $-$24.2320 & $3.149(1)$ & 3336 	& 0.320 & 6.627 & 2\\ 
300741820  	& 115.1874 	& $-$66.8087 & $3.168(1)$ & 2996 	& ...       & ...       & 1, 2\\  
272466980 	& 120.4676 	& $-$74.1727 & $3.318(1)$ & 3143 	& 0.397 & 5.802 & 2\\ 
389051009 	& 333.4603 	& $-$63.7028 & $3.436(1)$ & 2885 	& 0.120 & 9.228 & 1\\ 
453095489 	& 113.9139 	& $-$67.6945 & $3.699(1)$ & 3299 	& 0.285 & 6.488 & 1, 2\\ 
158596311 	& ~\,22.1261 	& $-$49.3526 & $3.711(1)$ & 3096 	& 0.281 & 6.525 & 2\\ 
441055710 	& 359.4175 	& $-$15.8517 & $3.722(1)$ & 2782 	& 0.140 & 8.541 & 2\\ 
220478846 	& ~\,75.8890 	& $-$56.5118 & $3.827(1)$ & 3665 	& 0.410 & 5.267 & 1, 2\\ 
350296377 	& ~\,83.6215 	& $-$54.2042 & $3.833(2)$ & 3410 	& 0.274 & 6.507 & 1\\ 
314888837 	& ~\,50.9265 	& $-$81.8043 & $3.920(1)$ & 2996 	& 0.149 & 8.214 & 1\\ 
~\,31381302 	& ~\,88.3600 	& $-$71.5638 & $3.965(1)$ & 2871 	& 0.109 & 9.069 & 1, 2\\ 
206502540 	& ~\,15.8985 	& $-$55.2656 & $3.998(1)$ & 2898 	& ...       & ...       & 2\\ 

\hline
\label{tbl:rotators}
\end{tabular}

{Notes: (a) {\ron Error bars in parentheses are in the last digit and error bars in the next digit are rounded to 1.} (b) The stellar temperature for each star is from TIC version 7 \citep{stassun18} and was obtained via the Mikulski Archive for Space Telescopes (MAST) online portal. (c) The masses and radii required to evaluate the fractional breakup rotation frequency are obtained from the absolute $K$ magnitudes and Eqns.~(\ref{eqn:MandR}). (d) The absolute $K$ magnitude is obtained from VizieR (\url{http://vizier.u-strasbg.fr/}; UCAC4 \citep{zacharias13} and the Gaia distance \citep{lindegren18}.  Gaia parallaxes are not available for three of the objects. (e) {\ron Sector 1 spans BJD-2457000 from 1325-1353, while Sector 2 covers 1354-1382.}
} 
\end{table*}

\begin{sidewaystable*}
\vspace*{250px}
\footnotesize
\begin{center}
\caption{Rapidly Rotating M Stars with Complex Rotation Modulation Profiles}
\begin{tabular}{lcccccccccc}
\hline
\hline
 Object Name			& 38820496 		& 177309964 		& 201789285 		& 206544316 		& 224283342 		& 234295610 		&  289840928  		& 332517282 		& 425933644 		& 425937691  		\\
\hline

RA J2000$^a$ 			& 7.19516			& 103.45258 		& 33.88868  		&  18.41884  		&  356.35758 		& 357.98325 		&  317.63150  		& 350.87860 		& 3.69942 		& 5.36556			\\
Dec J2000$^a$ 		& $-67.86237$		& -75.70396 		& $-56.45488$ 		& $-59.65974$ 		& $-40.33782$ 		& $-64.79293$ 		& $-27.18311$ 		& $-28.12114$ 		& $-60.06352$ 		& $-63.85226$		\\
$T_{\rm eff}$ [K]$^b$ 	& 2987			& 3289 			& ... 				& 3112 			& 3198 			& 3082 			& 3120 			& 3032 			& 3176 			& 2733 			\\  
d [pc] $^a$			& $44.1 \pm 0.1$	& $91.3 \pm 0.5$ 	& $45.4 \pm 0.2$  	& $43.1 \pm 0.1$ 	& $38.2 \pm 0.1$ 	& $48.2 \pm 0.1$	& $40.4 \pm 0.2$ 	& $39.1 \pm 0.1$ 	& $44.3 \pm 0.2$ 	& $43.9 \pm 0.2$ 	\\
pmra [mas/yr] $^a$		& $+89.2 \pm 0.1$	& $+4.7 \pm 0.1$ 	& $+95.3 \pm 0.1$ 	& $+96.7 \pm 0.1$ 	& $+108.4 \pm 0.1$ 	& $74.6 \pm 0.1$	& $+67.8 \pm 0.1$  	& $+109.3 \pm 0.1$ 	& $+88.8 \pm 0.2$ 	& $+91.6 \pm 0.2$ 	\\
pmdec [mas/yr]$^a$  	& $-50.9 \pm 0.1$	& $39.3 \pm 0.1$ 	& $-14.7 \pm 0.1$ 	& $-41.0 \pm 0.1$ 	& $-78.8 \pm 0.1$ 	& $-59.7 \pm 0.1$	& $-75.1 \pm 0.1$ 	& $-123.5 \pm 0.1$  	& $-56.9 \pm 0.2$ 	& $-53.9 \pm 0.2$	\\
G$^a$ 				& $14.77 \pm [2] $ 	& $14.45 \pm [2]$ 	& $15.63 \pm [2]$ 	& $12.95 \pm [2]$ 	& $13.66 \pm [1]$ 	& $13.89 \pm [1]$	 & $13.60 \pm [1]$ 	& $14.74 \pm [3]$ 	& $12.71 \pm [1]$ 	& $14.76 \pm [2]$ 	\\
G$_{\rm BP}$$^a$ 		& $16.80 \pm 0.01$ 	& $16.13 \pm 0.01$	 & $18.04 \pm 0.01$ 	& $14.61 \pm [6]$  	& $15.42 \pm [4]$ 	& $15.65 \pm [7]$	& $15.56 \pm [4]$ 	& $16.70 \pm 0.02$ 	& $14.31 \pm [6]$ 	& $17.12 \pm 0.02$ 	\\
G$_{\rm RP}$$^a$ 		& $13.42 \pm [4]$ 	& $13.19 \pm [6] $	 & $14.19 \pm [5]$ 	& $11.72 \pm [5]$ 	& $12.39 \pm [2]$ 	& $12.61 \pm [4]$	& $12.27 \pm [2]$ 	& $13.41 \pm 0.01$ 	& $11.48 \pm [3]$ 	& $13.34 \pm [7]$ 	\\
J$^c$				& $11.40 \pm 0.03$ 	& $11.36 \pm 0.03$ 	& $11.93 \pm 0.02$ 	& $9.95 \pm 0.02$ 	& $10.53 \pm 0.02$ 	& $10.72 \pm 0.026$	& $10.30 \pm 0.02$ 	& $11.41 \pm 002$  	& $9.71 \pm 0.02$ 	& $11.02 \pm 0.02$ 	\\
H$^c$ 				& $10.77 \pm 0.02$ 	& $10.73 \pm 0.03$ 	& $11.32 \pm 0.03$ 	& $9.34 \pm 0.03$ 	& $9.95 \pm 0.02$ 	& $10.17 \pm 0.024$	& $9.71 \pm 0.03$ 	& $10.83 \pm 003$ 	& $9.10 \pm 0.02$ 	& $10.48 \pm 0.03$ 	\\
K$^c$ 				& $10.5 \pm 0.02$ 	& $10.48 \pm 0.03$ 	& $10.95 \pm 0.02$ 	& $9.06 \pm 0.03$ 	& $9.69 \pm 0.02$ 	& $9.88  \pm 0.021$	& $9.41 \pm 0.02$ 	& $10.51 \pm 0.02$ 	& $8.83 \pm 0.02$ 	& $10.11 \pm 0.03$ 	\\
$M_K$$^{c,a}$ 		& 7.278			& 5.680 			& 7.666 			& 5.888 			& 6.784 			& 6.464			& 6.379 			& 7.545 			& 5.602 			& 6.896			\\
W1$^d$ 				& $10.31 \pm 0.02$ 	& $10.35 \pm 0.02$ 	& $10.66 \pm 0.02$ 	& $8.90 \pm 0.02$ 	& $9.50 \pm 0.02$ 	& $9.71 \pm 0.022$	& $9.06 \pm 0.02$ 	& $10.32 \pm 0.02$ 	& $8.70 \pm 0.02$ 	& $9.91 \pm 0.02$ 	\\
W2$^d$ 				& $10.09 \pm 0.02$ 	& $10.15 \pm 0.02$ 	& $10.40 \pm 0.02$ 	& $8.71 \pm 0.02$ 	& $9.30 \pm 002$ 	& $9.51 \pm 0.019$	& $8.81 \pm 0.02$ 	& $10.13 \pm 0.02$ 	& $8.51 \pm 0.02$ 	& $9.66 \pm 0.02$ 	\\
W3$^d$ 				& $9.95 \pm 0.05$ 	& $10.00 \pm 0.04$ 	& $10.16 \pm 0.05$ 	& $8.58 \pm 0.02$ 	& $9.13 \pm 0.03$ 	& $9.27 \pm 0.034$	& $8.77 \pm 0.03$ 	& $9.90 \pm 0.05$ 	& $8.36 \pm 0.02$ 	& $9.40 \pm 0.03$ 	\\
W4$^d$				& $\gtrsim 8.65$ 	& $\gtrsim 9.35$ 	& $\gtrsim 8.5$ 		& $8.46 \pm 0.26$ 	& $\gtrsim 8.5$ 		& $\gtrsim 8.46$	& $8.41 \pm 0.33$ 	& $\gtrsim 8.5$ 		& $8.17 \pm 0.25$ 	& $\gtrsim 8.5$		\\
Radius [$R_\odot$]$^e$ 	& $0.28 \pm 0.01$	& $0.49 \pm 0.02$ 	& $0.24 \pm 0.01$ 	& $0.46 \pm 0.01$ 	& $0.34 \pm 0.01$ 	& $0.38 \pm 0.01$	& $0.38 \pm 0.01$ 	& $0.25 \pm 0.01$  	& $0.51 \pm 0.02$ 	& $0.32 \pm 0.01$ 	\\
Mass [$M_\odot$]$^e$ 	& $0.25 \pm 0.01$	& $0.49 \pm 0.02$ 	& $0.18 \pm 0.01$ 	& $0.35 \pm 0.01$ 	& $0.25 \pm 0.01$ 	& $0.37 \pm 0.01$	& $0.29 \pm 0.01$ 	& $0.19 \pm 0.01$ 	& $0.38 \pm 0.01$ 	& $0.30 \pm 0.01$ 	\\
Period [hr]$^f$ 			& $15.727(6)$		& $10.881(6)$ 		& $3.638(1)  $		& $7.727(3)$     	& $21.298(11)$ 	& $18.288(13)$		& $4.780(3)$ 		& $9.660(9)$		& $11.669(6) $		& $4.816(2)$		\\
Association$^f$ 		& THA			& CAR 			& THA/CAR		& ABDMG 		& COL 			& THA			& BPMG 			& ABDMG 		& THA 			& THA			\\
{\ron Data sectors$^g$} & 1, 2 & 1, 2 & 2 & 1, 2 & 2 & 1 & 1 & 2 & 1, 2 & 1, 2\\
\hline
\label{tbl:weird}
\end{tabular}
\end{center}

{Notes.  {\ron Error bars in parentheses are in the last digit and error bars in square brackets are in the following digit.}  (a) Gaia DR2 \citep{lindegren18}.  (b) TIC v.7 \citep{stassun18}. (c) 2MASS catalog \citep{skrutskie06}.  (d) WISE point source catalog \citep{cutri13}.  (e) See Eqns.~(\ref{eqn:MandR}); \citet{mann16,mann18}. (f) See Table 3 for more details \citet{gagne18}.} (g) {\ron Sector 1 spans BJD-2457000 from 1325-1353, while Sector 2 covers 1354-1382.}
\end{sidewaystable*}

\section{{\em TESS} Observations}
\label{sec:observations}
In this study we utilized the 2-minute cadence (short cadence) data collected by {\em TESS} during Sectors 1 and 2.  The targets selected for short cadence are chosen from the candidate target list (CTL) \citep{stassun18}. The light curves were produced by the Science Processing Operations Center (SPOC) pipeline \citep{jenkins16}, which is operated by NASA Ames Research Center, and are available online\footnote{\url{http://archive.stsci.edu/tess/bulk_downloads.html}}. For Sectors 1 and 2, 2-minute cadence data were obtained for a total of 24,809 unique targets of which 7,074 were observed in both sectors. Although the primary goal of {\em TESS} is to search for terrestrial planets that transit nearby bright stars, the large number of targets monitored with 2-minute cadence enables the study of other astrophysical phenomena on time scales down to the Nyquist limit of 360 cycles d$^{-1}$ (4-minute period).

We opted to use the SAP\_FLUX light curve data {\ron and do our own detrending}. For each light curve, we split the non-detrended SAP\_FLUX time series into separate orbits and then normalized the data for each orbit by the mean flux of that orbit.  Since we were primarily interested in finding rapidly rotating M dwarfs, we detrended each orbit using a 3-day spline filter to remove any longer-timescale fluctuations in the flux.   

We note that many of the M dwarfs in this study were added to the CTL as a part of the Cool Dwarf specially curated catalog \citep{muirhead18}.

\vspace{0.3cm}

\section{Analysis}
\label{sec:methods}

\subsection{Fourier Analysis}

We used a basic fast Fourier transform (FFT) method to search for rapidly rotating M dwarf stars. The FFT method is fast and efficient when used to search for periodic signals that have high-duty cycles, or, equivalently, smoothly varying rotation profiles. The FFT search used here is similar to the one employed by \citet{sanchis-ojeda13, sanchis-ojeda14} but is optimized for the {\em TESS} data.  Given that the data set essentially consists of samples uniformly spaced in time and minimally affected by window functions, we did not see that any particular advantage would be gained by using the Lomb-Scargle transform \citep{lomb76, scargle82}. {\ron Accurate periods and the corresponding uncertainties for any interesting objects are found via a Stellingwerf transform \citep{stellingwerf78} which utilizes a phase dispersion minimization algorithm.}

The preprocessed data for each orbit are stitched together into one data array, the mean flux is subtracted, and the data gaps are filled with zeroes.  The data array was extended by padding with zeroes to a power of 2 number of bins which represents approximately 8 times the length of the non-zero part of the data array.  This yields Fourier `interpolation' in frequency.  In the case where only 1 {\em TESS} Sector (2 satellite orbits) of data ($\sim$20,000 data points) is available, the final data array comprises 2$^{17}$ bins. We then compute the FFT using Python with the SciPy FFT package, which takes $\approx 0.1$ second for each object on a computer with a 2.9 GHz Intel Core i7 Processor.  From the complex FFT we form the amplitude spectrum. 

We apply the data preparation and Fourier transform operations to the data for all of the 2-minute cadence target stars.  However, in this paper we focus on the results of our search for rapidly rotating M stars.

\subsection{Finding rapidly rotating M stars}
\label{sec:rapidMstars}

To find rapidly rotating M dwarfs, we first select all stars among the 2-minute cadence targets with reported stellar temperatures $T_{\rm eff} \lesssim$ 4000 K.  Rapidly rotating M dwarfs are known to be active flaring stars; flares can add significant noise to the Fourier transform, and thereby adversely affect the search (see Sect.~\ref{sec:methods}). Therefore flares are removed prior to application of the FFT using a 5-$\sigma$ flare detection threshold and flare fitting technique developed and discussed in \citet{gunther19}.  

To detect rapidly rotating M-dwarfs, we compute a local, i.e., 128 frequency bin, mean and local rms ($\sigma$) for each bin in each Fourier amplitude spectrum.  We consider a detection to be significant if at least one value in the amplitude spectrum exceeds the local mean by at least 5 local $\sigma$ anywhere in the frequency range\footnote{As noted earlier, the {\em TESS} short-cadence frequency range reaches 360 cycles d$^{-1}$; however, we truncated our search at 48 cycles d$^{-1}$ because even an M star of mass 0.1 $M_\odot$ and radius 0.15 $R_\odot$ cannot rotate with a period below 1/2 hr without breaking up due to centrifugal forces (see discussion in Sect.~\ref{sec:collection} and Eqns.~\ref{eqn:MandR}).} of 1 - 48 cycles d$^{-1}$, {\em and}, is accompanied by at least one higher harmonic or subharmonic that exceeds its local mean by 3 local $\sigma$. We vetted the targets that pass this threshold by eye to confirm detection, and, in the process, eliminated eclipsing M-star binaries and objects with other obvious non-rotational signals.  The result was the detection of approximately 500 potentially rapidly rotating M dwarfs with periods shorter than 1 day in either one or both sectors, which accounts for roughly 1/10th of the M dwarf stars in the CTL for Sectors 1 and 2.

\vspace{0.3cm}

\subsection{The collection of rapidly rotating M stars}
\label{sec:collection}

\begin{figure}
\begin{center}
\includegraphics[width=\columnwidth]{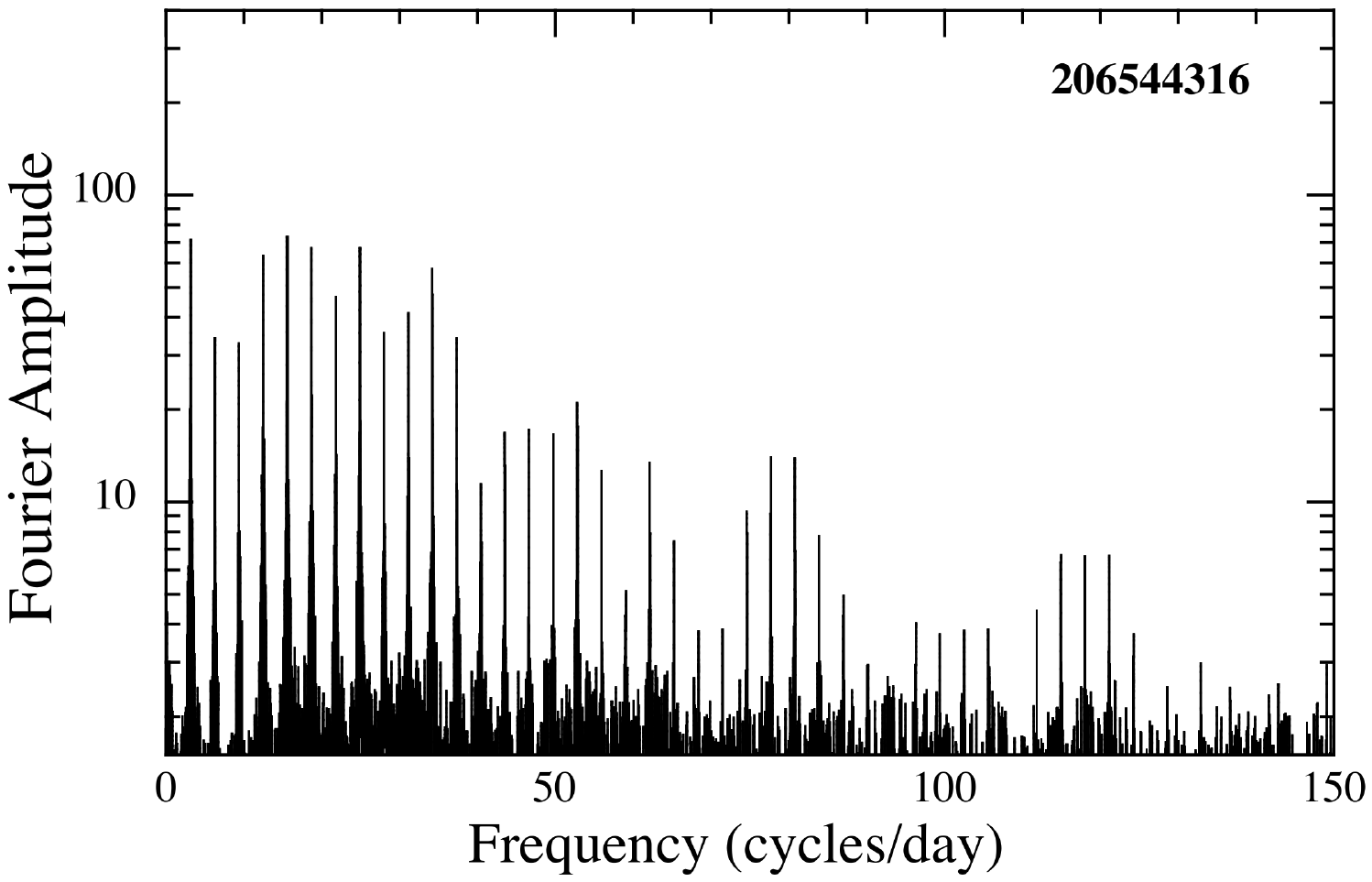}
\caption{Fourier transform of the Sector 2 {\em TESS} data for TIC 206544316.  The frequencies corresponding to the rotation period of 7.7274 hours and more than 40 detectable harmonics are evident, indicating there is considerable high-frequency structure in the light curve down to timescales of $\sim$4 min.  }
\end{center}
\label{fig:fft} 
\end{figure}

\begin{figure}
\begin{center}
\includegraphics[width=\columnwidth]{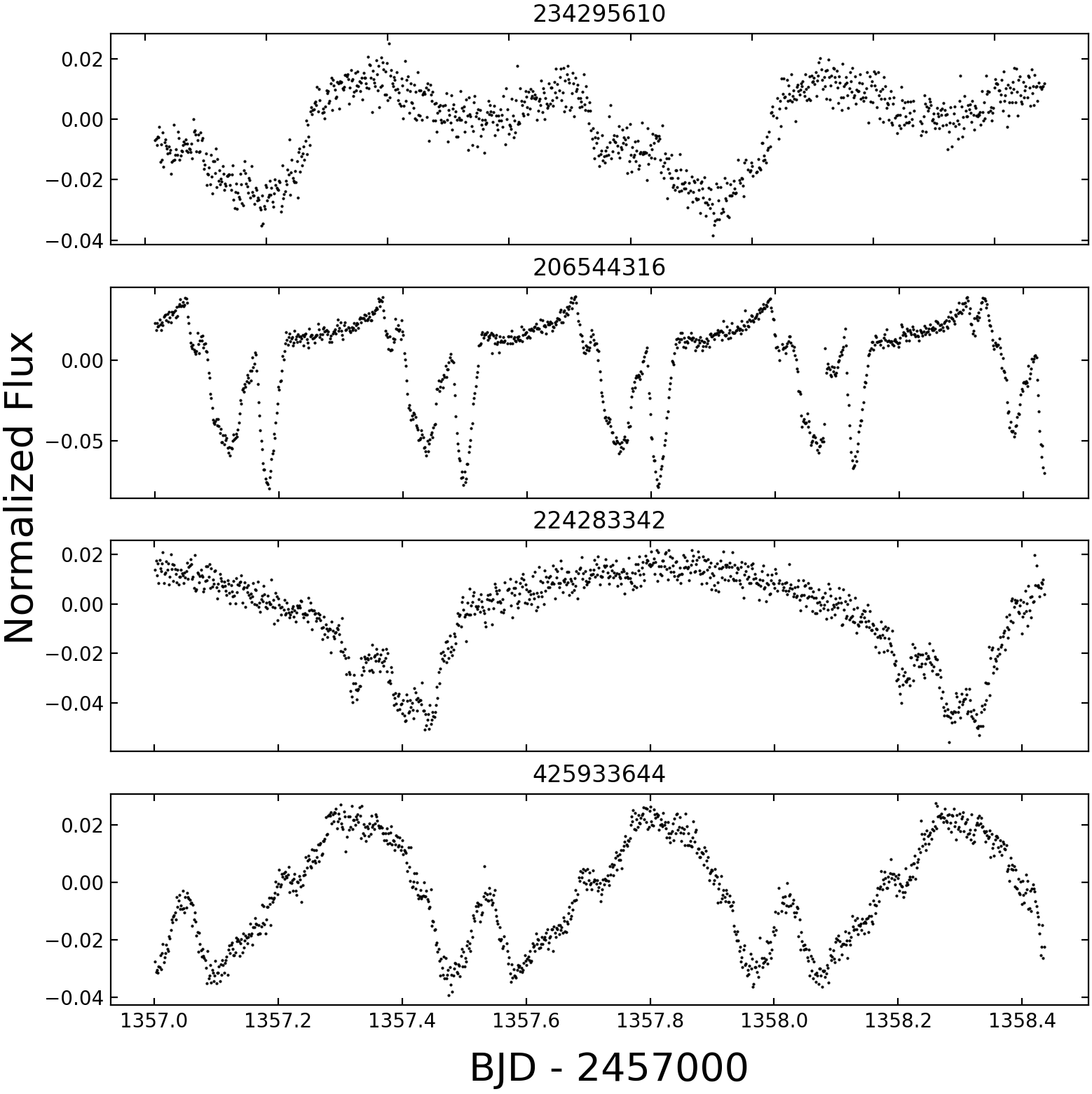}
\caption{Segments of raw data for four illustrative light curves of the rapid M-star rotators we have found in the {\em TESS} Sectors 1 \& 2 data that exhibit highly structured rotational modulations.  Each plot shares the same time axis and is 1.4 days long.}
\label{fig:rawLC}
\end{center}
\end{figure} 

\begin{figure*}
\begin{center}
\includegraphics[width=0.99\textwidth]{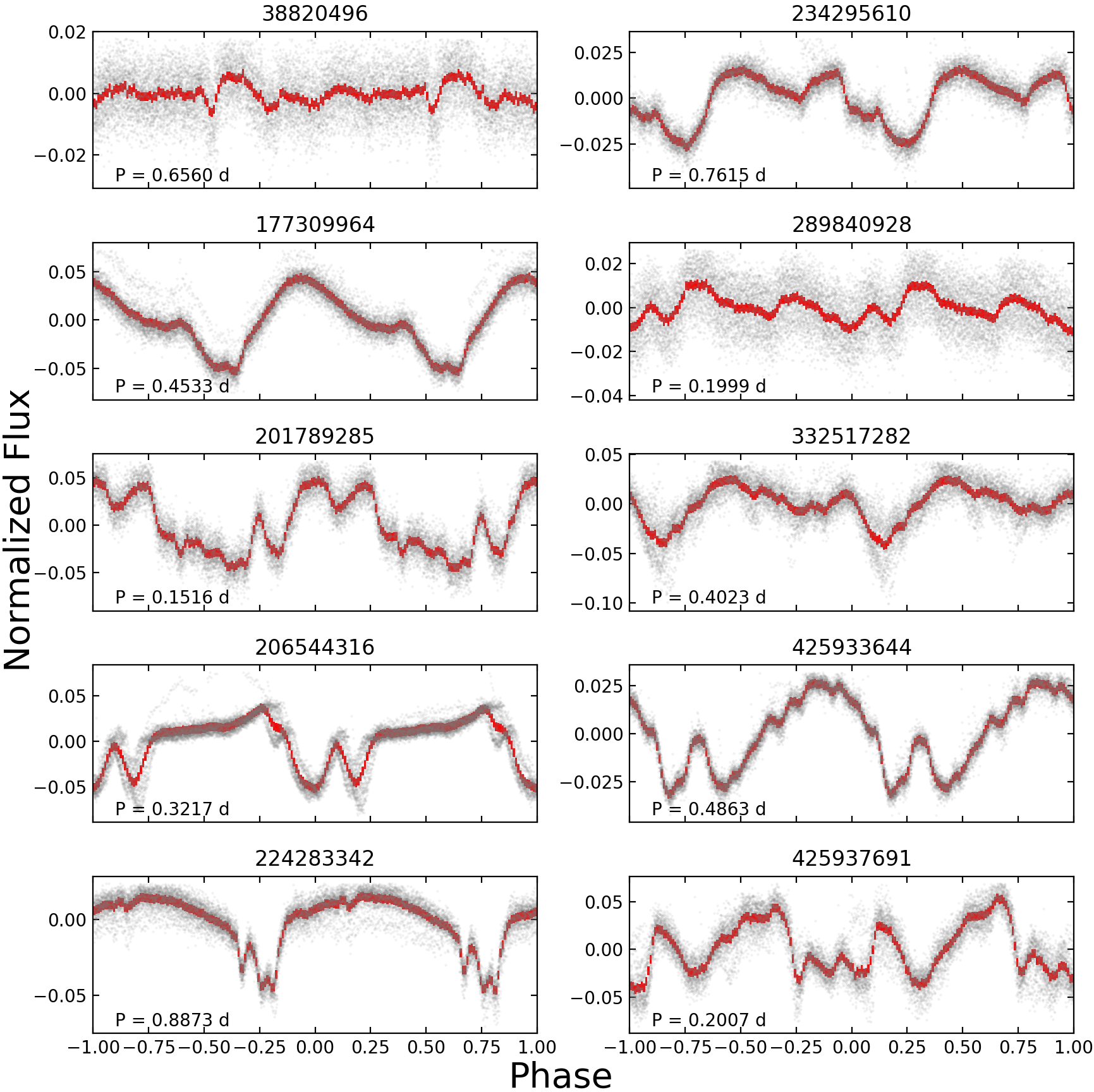}
\caption{Folded light curves for the 10 rapidly rotating M stars with complex rotational modulation profiles found in Sectors 1 and 2 with {\em TESS}. The fold periods are indicated.  The red points are the folded and binned data, while the gray points are individual flux measurements (with flares removed). Almost all of the high-frequency features seen in the folds are statistically significant. The large fluctuations in the data for TIC 289840928 are due to the presence of another independent period at 15.606 hr.  By contrast, the data for TIC 38820496 are intrinsically noisy.}
\label{fig:folds} 
\end{center}
\end{figure*}

We found a total of {\NSecTotalFastMdwarf} M-dwarf stars to be rotating with periods less than 4 hours among {\NSecTotalRRMdwarf} M-dwarfs with periods below 1 day and $\sim$1000 with periods below 2 days. With a period of 1.63 hr, TIC 204294226 is the most rapidly rotating M dwarf discovered thus far.  

The {\NSecTotalFastMdwarf} fastest rotators are listed in Table \ref{tbl:rotators} along with coordinates and effective temperatures.  The ratio of rotation frequency to breakup angular frequency is also listed for each star.  The estimated breakup frequency of each star is based on the stellar mass, $M_*$, and radius, $R_*$, expressed as simple functions of the absolute K magnitude, $M_K$, following \citet{mann16, mann18}:
\begin{eqnarray}
R_*  & \simeq & 1.9515 - 0.3520 M_K + 0.01680 M_K^2 \label{eqn:MandR} \\
\log_{10}(M_*) & \simeq & a + b M'_K+c M'^2_K+d M'^3_K+e M'^4_K+f M'^5_K \nonumber
\end{eqnarray} 
where $M'_K \equiv M_K - 7.5$, and the coefficients $a-f$ are: \\
$a =-0.642$,
$b=-0.208$,
$c=-0.000843$,
$d=0.00787$,
$e=0.000142$, and
$f=-0.000213$.  The ratio of $\Omega/\Omega_{\rm crit}$ is computed from the expression:
\begin{equation}
\frac{\Omega}{\Omega_{\rm crit}} =\frac{2 \pi}{P_*}\sqrt{\frac{R_*^3}{G M_*}} \simeq  2.8 \sqrt{\frac{r_*^3}{m_*}} \left(\frac{{\rm hr}}{P_*}\right)
\label{eqn:omega}
\end{equation}
where $m_*$ and $r_*$ are the mass and radius in solar units and $P_*$ is the rotational period {\ron in hours}.

\subsection{Identifying highly structured rotation profiles}

In the process of vetting the rapidly rotating M stars, we discovered a number of stars with highly structured rotational modulation patterns.  The first clue came from examining the unusual harmonic structure in their amplitude spectra.  An extreme example of this is shown in Fig.~\ref{fig:fft}. While the amplitude spectrum of a rotating M star usually contains 2-4 detectable harmonics, TIC 206544316 exhibits 40 higher harmonics.  This is indicative of a highly structured rotation profile.  

The raw light curves for four of these objects are shown, as examples, in Fig.~\ref{fig:rawLC}. Note that each of these has sharp structure that occurs over small fractions of a rotation cycle.  

In all, we find 10 similarly structured rotation curves among our sample of rapid rotators.  Each of these objects exhibits 10 to 40 significant harmonics in the frequency range 1--96 cycles/day. These contrast with most rapidly rotating M dwarfs that have comparatively simple modulation patterns and therefore exhibit only 1--5 significant harmonics. The folded light curves for these ten rotators with complex modulation patterns are presented in Fig.~\ref{fig:folds} and some of their properties are presented in Table \ref{tbl:weird}. 

We see in Figure \ref{fig:folds} an array of exotic rotational modulation patterns.  To the extent we can generalize, the amplitudes of the modulations are typically 5\% to 10\% peak to peak.  The amplitudes and structures of the profiles seem roughly independent of rotation period between 2.5 and 21 hours.  The structures of the profiles are not simple to describe; the term ``scalloped shell'' \citep{stauffer17}, while descriptive, does only partial justice. `Highly structured', `comprised of many Fourier components', and `multi- and sharply-peaked' further describe the profile shapes.  

\subsection{Variability of the rotation profiles over time}

\begin{figure}
\centering
\includegraphics[width=1.0\columnwidth]{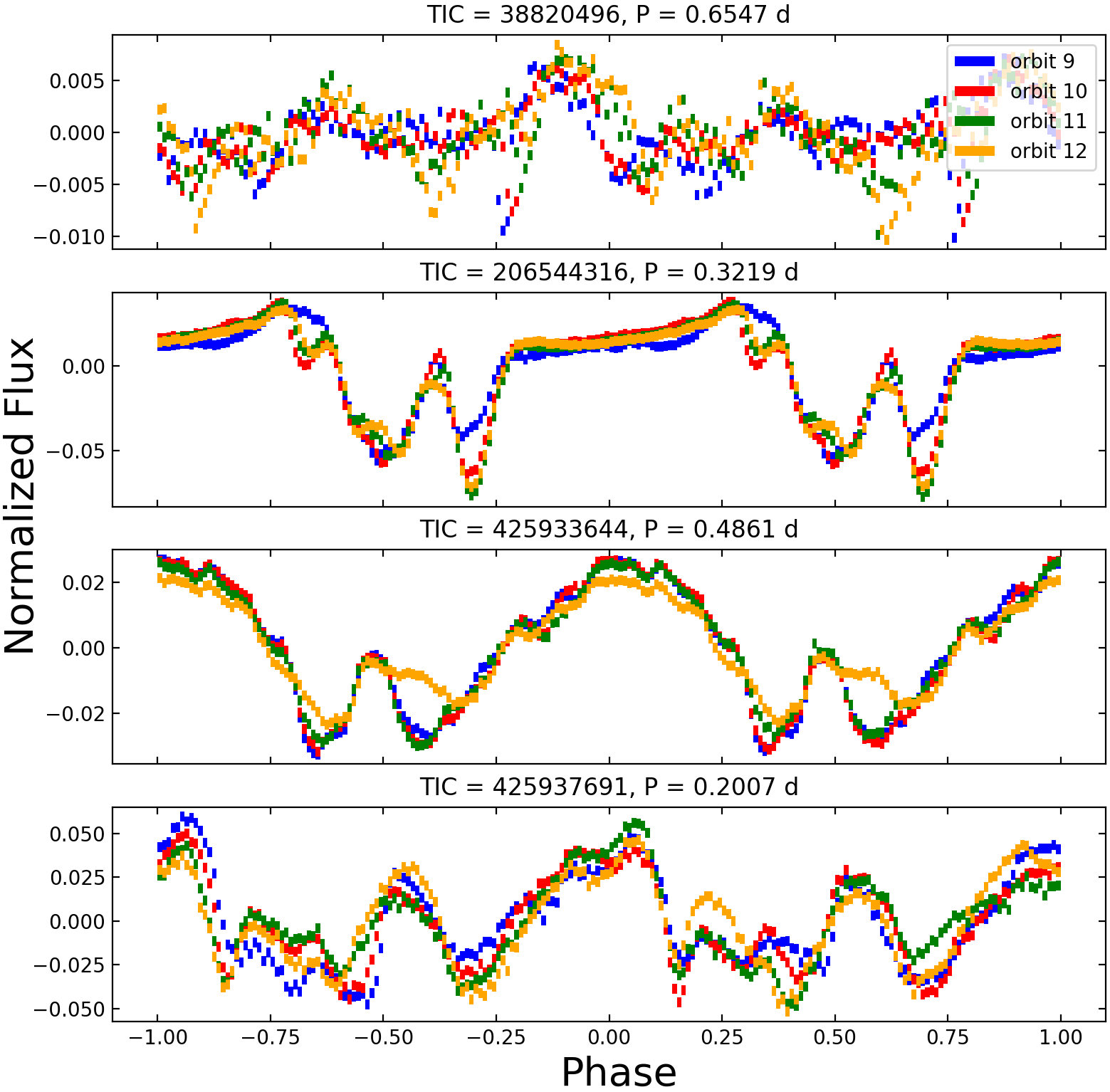}
\caption{Four examples of persistence of the rotational modulation patterns from {\em TESS} orbit to orbit. Orbits 9 and 10 (color coded blue and red) comprise {\em TESS} Sector 1 data, and similarly orbits 11 and 12 (green and orange) represent Sector 2 data.}
\label{fig:changes}  
\end{figure}

Fig.~\ref{fig:changes} shows the light curves of four of our highly structured rotators for which data was obtained from both Sectors 1 \& 2, i.e., from {\em TESS} orbits 9, 10, 11, and 12\footnote{Although there is data for TIC 177309964 from both sectors, they are not plotted because the modulation pattern is quite stable over the course of two months.}. Each color-coded folded profile represents the data taken in one {\em TESS} orbit, i.e., two weeks. While there are clearly variations in the detailed shapes of the rotational modulations, the basic profiles remain recognizably similar over two months.  

\begin{figure}
\begin{center}
\includegraphics[width=\columnwidth]{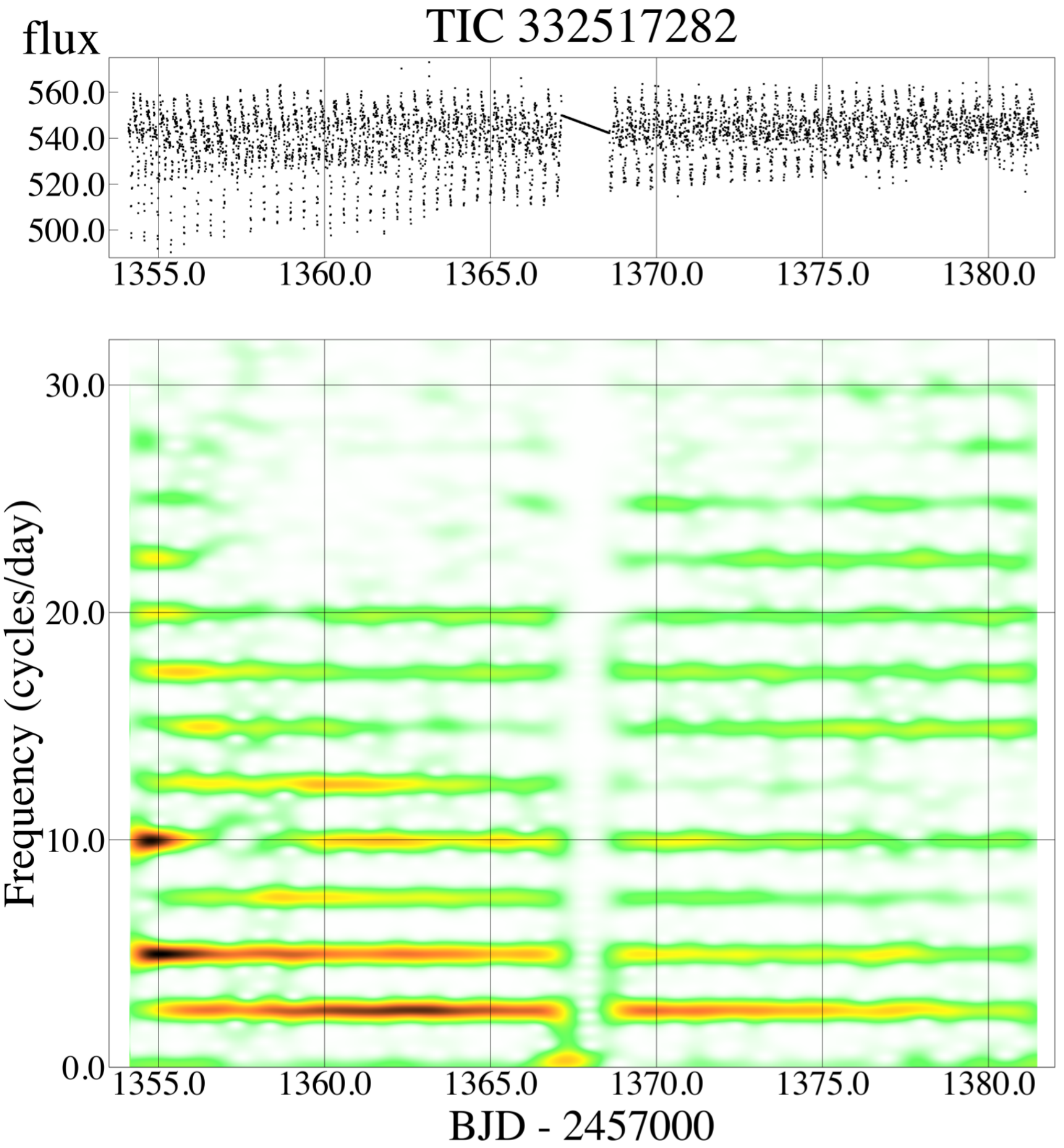}
\caption{Time-evolving Fourier amplitude spectrum for TIC 332517282 showing the stability of the modulation frequencies. The top panel shows the corresponding time series {\ron during Sector 2} .  The full width at half-maximum (FWHM) of the Gaussian window is 0.94 days and the data are binned to 7.2 min, leading to a frequency resolution and Nyquist frequency of 1.1 cycles d$^{-1}$ and 200 cycles d$^{-1}$, respectively.  The Gaussian window is stepped by $\sim$0.82 hours between Fourier transforms.}
\label{fig:sonogram} 
\end{center}
\end{figure}

In Fig.~\ref{fig:sonogram} we show the time evolution of the Fourier amplitude spectrum for TIC 332517282.  For these calculations we used the software package {\tt TiFrAn} (for details see \citealt{rappaport14}). Even though the amplitude of some of the features in the rotational modulation pattern are clearly changing with time, the fundamental frequency and its numerous harmonics remain essentially constant to within the frequency resolution over the course of a month.

\subsection{Comparison with highly structured rotators found in Upper Sco and $\rho$ Oph} 

The rotational modulation profiles of the 10 M stars discussed in this work are very similar to those that \citet{stauffer17} describe as ``scalloped shell'' modulation profiles.  Of course, the 2-minute time resolution of the {\em TESS} data, when compared to the 30-minute resolution of the {\em K2} data, allows for higher-frequency structure to be recorded.  The `scalloped shell' rotators in Upper Sco and $\rho$ Oph have a typical age of 5-10 Myr, a typical distance of 135 pc, and a median $K_s$ magnitude of 10.5 (median $M_K \simeq 4.9$).  The comparable rotators in this work are $\simeq 45$ pc away (closer), are $\simeq 45$ Myr old (older), and have a similar median $K_s$ magnitude, but a median $M_K \simeq 6.4$, which is notably 1.5 magnitudes less luminous than the `scalloped shell' modulators in Upper Sco and $\rho$ Oph.  We also note that the range of periods of our 10 systems of 2.5 to 21 hours is comparable to the range of 6.2 to 15.4 hr of the `scalloped shell' modulators in Upper Sco and $\rho$ Oph.  Finally, we point out that the WISE W3 fluxes of the stars in one set are similar to those of the other set, and the upper limits in the W4 band for the two sets of stars are similar.

\section{Spectroscopic Observations}
\label{sec:spectra}

\begin{figure}
\centering
\includegraphics[width=1.0\columnwidth]{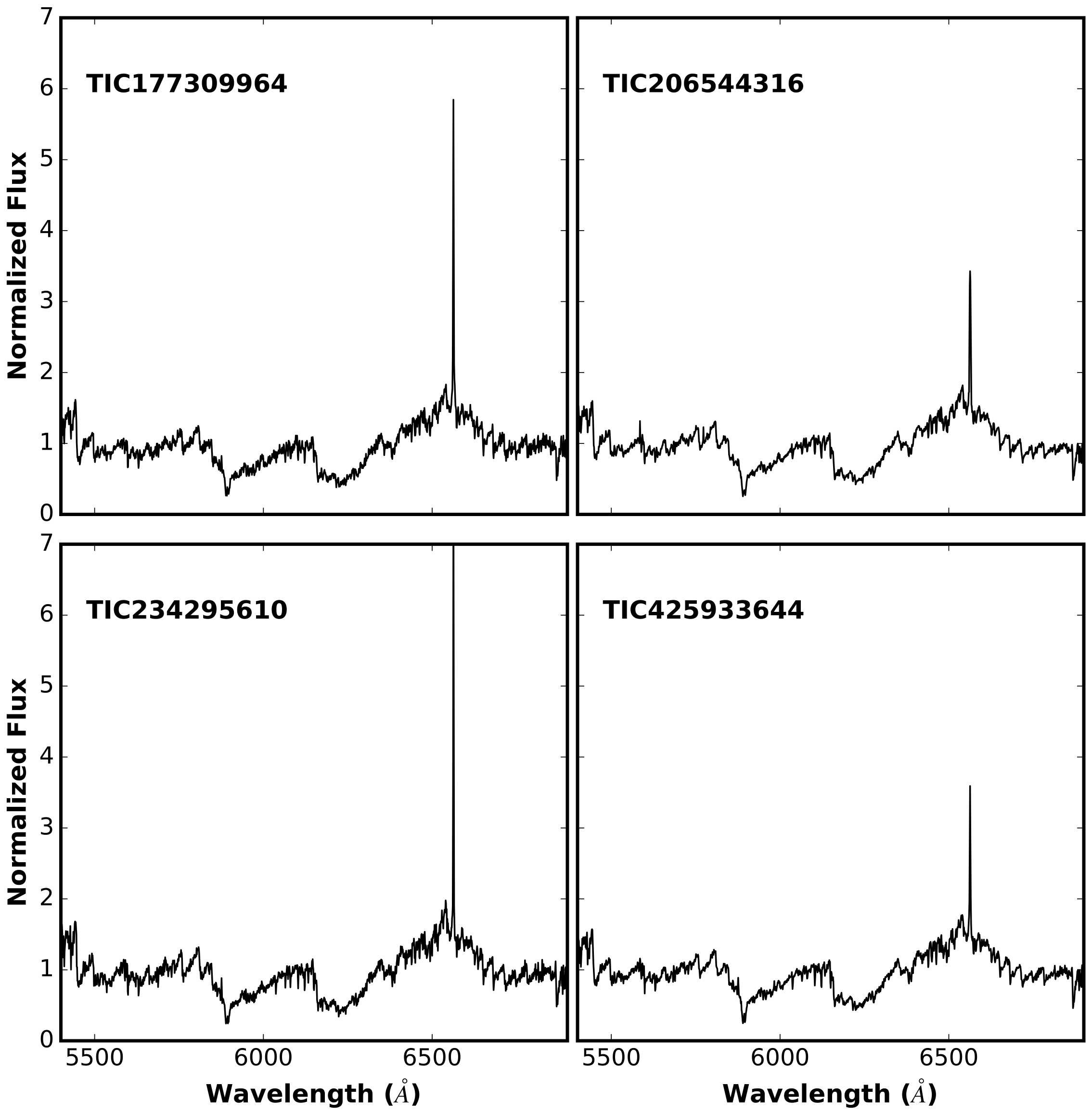}
\caption{Low resolution spectra for four of the rapidly rotating M-dwarfs obtained with the ANU 2.3-m WiFeS facility. In each spectrum, we see strong H$\alpha$ emission features at equivalent widths of 4-8\,\AA, indicative of strong chromospheric activity. We also find no evidence of binarity in any of the spectra.  {\ron For comparability, each spectrum is normalized by the mean flux between 6000-6200 \AA.}}
\label{fig:WiFeS} 
\end{figure}

We obtained low and medium resolution spectra for a select sample of the rapidly rotating M-dwarfs with complex rotational modulations to search for activity signatures and additional signs of rapid rotation. Low resolution observations were obtained for four of the systems using the Wide Field Spectrograph \citep[WiFeS, ][]{dopita07} on the ANU 2.3-m telescope at Siding Spring Observatory, Australia, on the nights of 2019 January 18 and 19. The data covered the wavelength region of $5200-7000$\,\AA\ at a resolution of $R \equiv \lambda / \Delta \lambda = 7000$, and were reduced per the procedure described in \citet{bayliss13}. The spectra are displayed in Figure~\ref{fig:WiFeS}. The spectra of all of the M-dwarfs sampled by WiFeS exhibit strong H$\alpha$ emission features with equivalent widths in the range $4-6$\,\AA, and none shows signs of the composite nature expected if a system is composed of multiple stars. These H$\alpha$ equivalent widths are quite typical of rapidly rotating M stars.

\begin{figure}
\begin{center}
\includegraphics[width=0.99 \columnwidth]{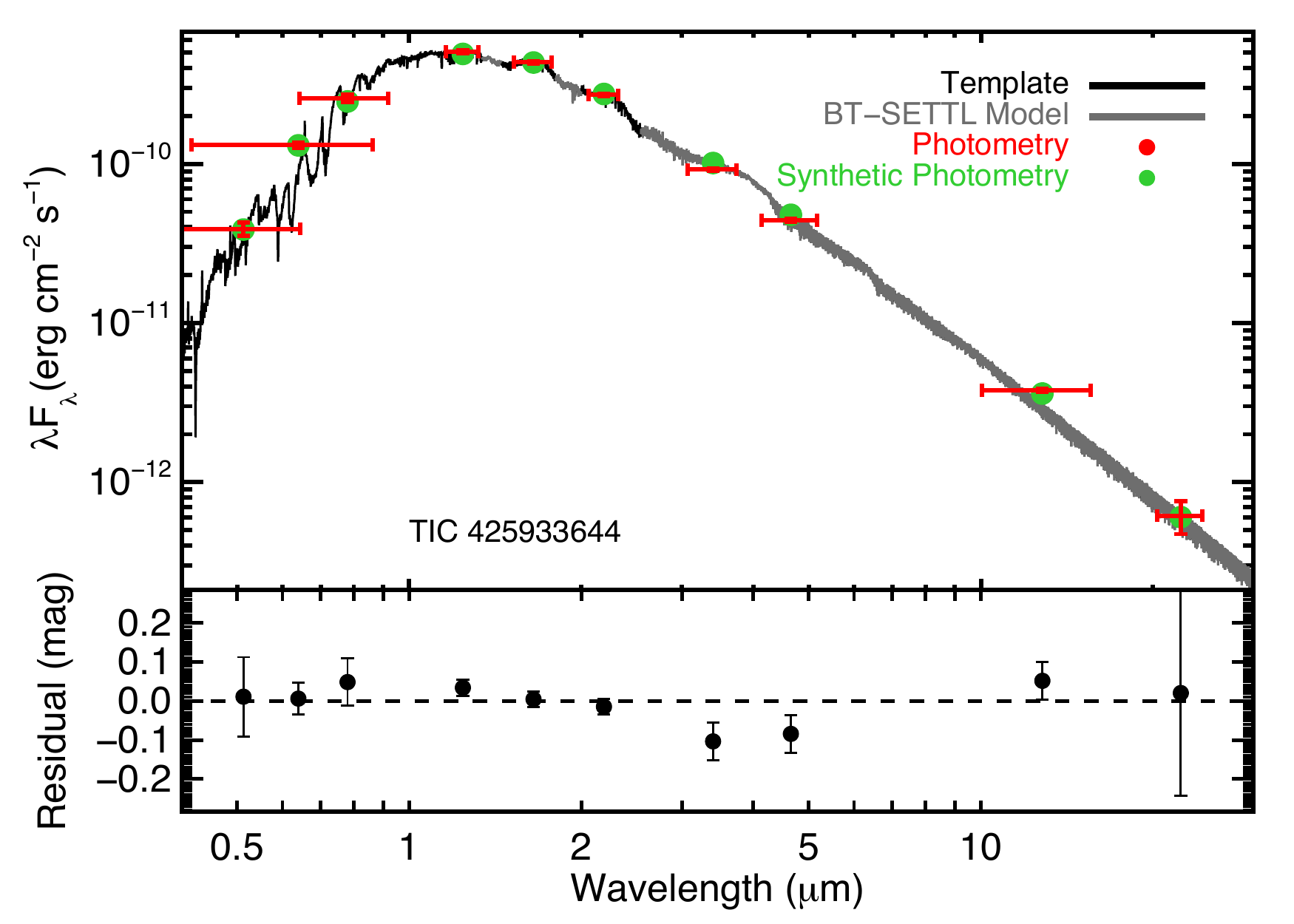}
\includegraphics[width=0.99 \columnwidth]{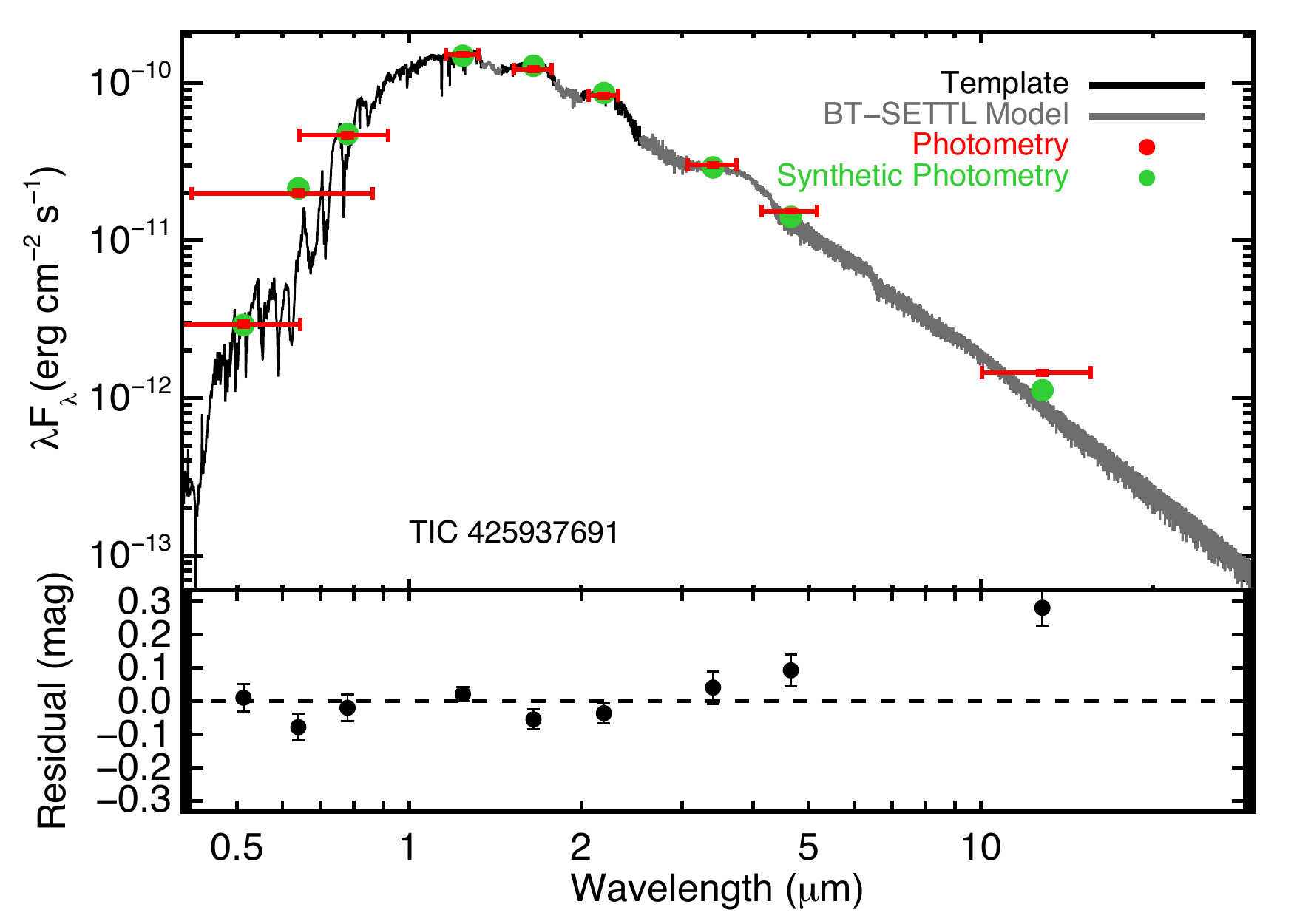}  
\includegraphics[width=0.99 \columnwidth]{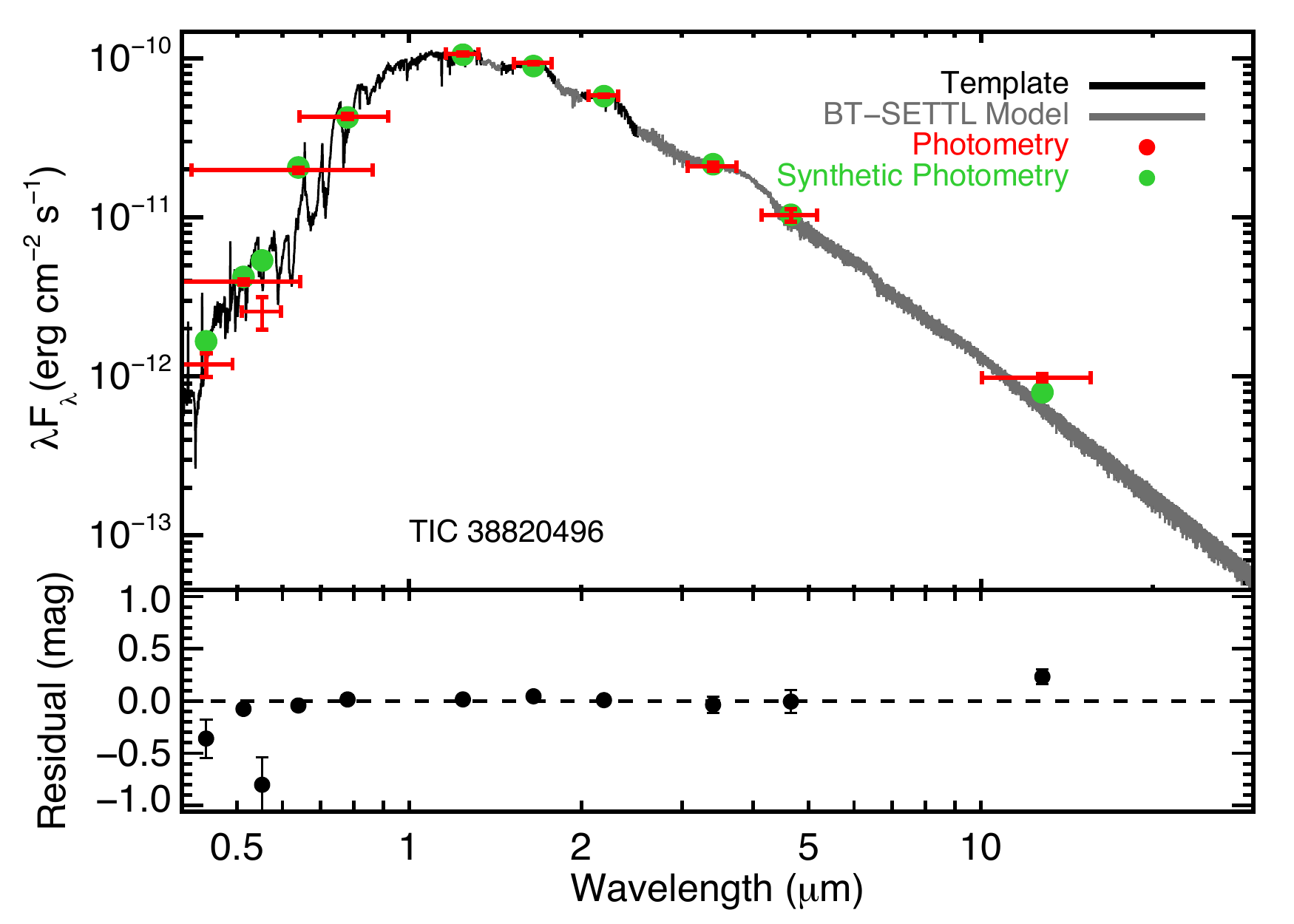}
\caption{Spectral energy distributions of TIC 425933644, TIC 425937691, and TIC 332517282, with the optical defined by the Gaia $G_\mathrm{BP}$ and $G_\mathrm{RP}$, and the infrared by the 2MASS and WISE magnitudes. Observed magnitudes are shown in red, with vertical errors corresponding to the observed magnitude uncertainties (including stellar variability) and the horizontal errors corresponding to the width of the filter. Synthetic magnitudes from the best-fit template spectrum (black) and BT-SETTL model (grey) are shown as green circles.}
\label{fig:sed}
\end{center}
\end{figure}  

In addition to the low resolution spectra from WiFeS, we also obtained a medium resolution spectrum of TIC 206544316 using the echelle spectrograph on the ANU 2.3 m telescope. The echelle yields a spectral resolution of $R=24000$ over the wavelength region $4200-6725$\,\AA, and the observations were reduced as per \citet{zhou14}. The echelle spectrum of TIC206544316 revealed that it is rapidly rotating at a rate of $v_\mathrm{eq}\sin i \sim 77\,\mathrm{km\,s}^{-1}$, consistent with the equatorial rotational velocity of the star derived from the photometric period. As with the WiFeS spectra, no obvious signatures of spectral blending indicative of multiple stars in the system are apparent. 

{\ron We find no evidence of ongoing accretion in these M-dwarf systems. The H-$\alpha$ emission strengths as measured from our spectra are in-line with that expected from rapidly rotating field M-dwarfs \citep{newton16}, and as such are not indicative of accretion. It is difficult to quantitatively constrain the extent of veiling in our low-resolution spectra.  We do not have accurate measurements of the rotational broadening, and as such it is more difficult to match the low-resolution spectra against synthetic templates while simultaneously fitting for line dilutions due to veiling.  We suggest future observations at high resolution that may be able to address this problem.}

To further probe the spectra of our ten highly structured rotators, especially in the NIR, we utilized the observed spectral energy distribution (SED). We made use of photometric magnitudes from Gaia $G_\mathrm{BP}$ and $G_\mathrm{RP}$ \citep{gaia18}, 2MASS $J$, $H$, $Ks$ bands \citep{cutri03}, and WISE $W1$, $W2$, $W3$, and $W4$ where available \citep{cutri13} to describe the SED of all ten of our stars listed in Table \ref{tbl:weird}. 

We compared the SEDs to a set of un-reddened optical and near-infrared spectra from the TW Hydra or $\beta$ Pic moving groups \citep[10-25\,Myr,][]{bell15}, following the method from \citep{mann16b}, which we briefly summarize here. We restricted our comparison to templates with spectral types or $T_{\rm{eff}}$ estimates consistent with the stars in Table \ref{tbl:weird}. Spectral types and $T_{\rm{eff}}$ values for the SED-fitting templates were determined in an identical manner, and hence were on the same overall scale. Gaps in the coverage of each template were filled with PHOENIX BT-SETTL models \citep{allard12}. We assumed zero reddening, as all targets are within the local bubble \citep{Sfeir99}. 

After fitting all ten SEDs we reach several conclusions.  (1) Four of the SEDs show no NIR excess, e.g., TIC 206544316 and TIC 425933644 (see top panel of Fig.~\ref{fig:sed}) have W4 magnitudes, and both are in excellent agreement with the model+template. TIC 38820496 and TIC 177309964 are well fit without any need for an excess. (2) Three of the systems show a clear NIR excess, e.g., TIC 425937691 (see middle panel of Fig.~\ref{fig:sed}) and TIC 234295610 have W3 magnitudes that are  $\gtrsim 4 \, \sigma$ above the model, even considering the range of possible model fits. TIC 201789285 is also a fairly clear case, with a possible W2 excess and a larger break at W3.  (3) The remaining three systems fall in the ambiguous range, e.g., TIC 224283342 appears to show a W3 excess, but it is only barely significant. TIC 332517282's W3 excess is not significant, just suggestive (see bottom panel of Fig.~\ref{fig:sed}). 

These results may be similar to what we would find if we were to randomly sample stars in these young stellar associations. But, the fact that some show clear excesses does lend credence to the notion that dust is involved. In particular, these excesses are consistent with what we would expect to see from a small dust disk near the star whose projected area is less than an order of magnitude larger than that of the host star, and having equilibrium grain temperatures of $\lesssim 500$ K (see Sect.~\ref{sec:ring}).

\section{Modeling the Complex Lightcurves with Spots}
\label{sec:spots}

\subsection{Fitting stellar spots with {\scshape allesfitter}}
We studied whether the periodic variability of the 10 M stars may be explained by the presence of spots on the stellar surfaces.  For this, we use the {\scshape allesfitter} package {\guentherinprep} to fit spot models of different complexities.  Using Nested Sampling \citep{skilling2006}, we explore the entire physical parameter space, and compute the Bayesian evidence for different models.

\begin{figure*}[!ht]
\centering
\includegraphics[width=0.8\textwidth]{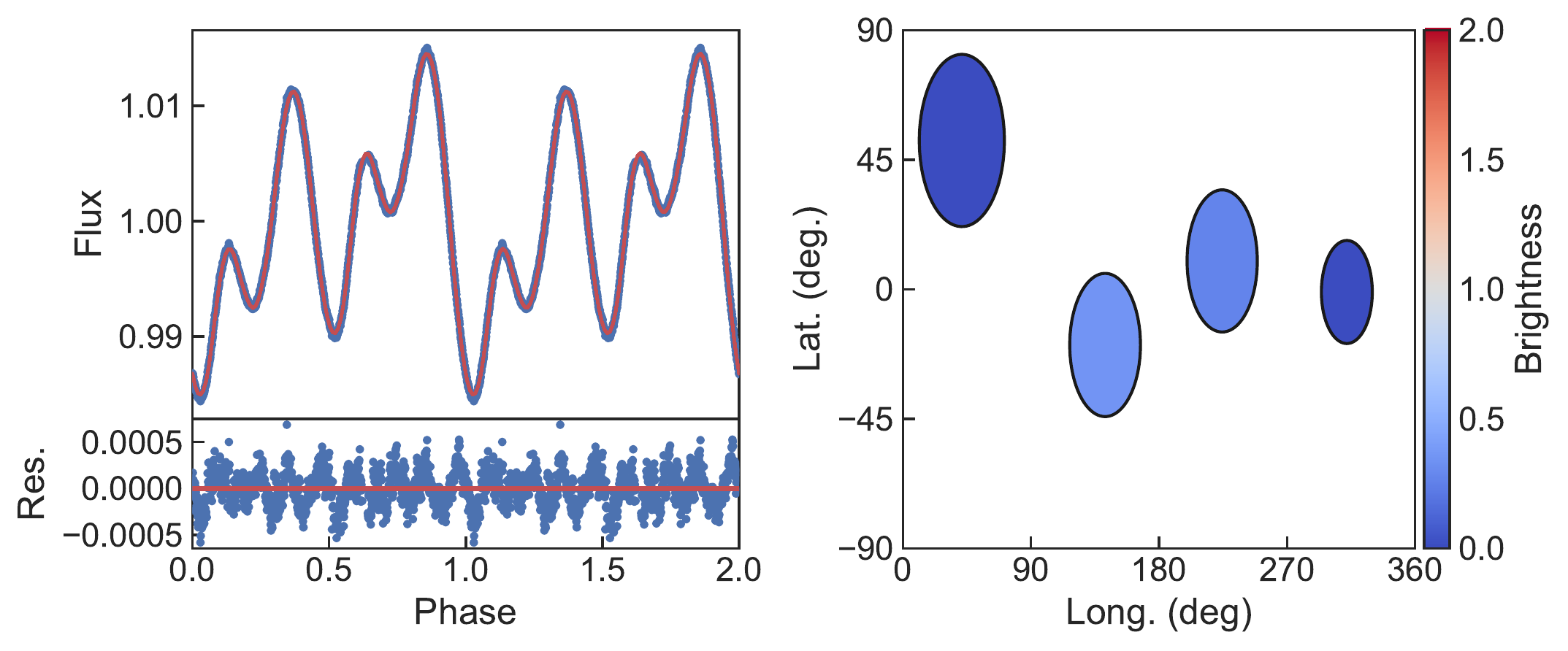}
\caption{A model of four spots on one star can describe nearly all of the periodic variability of HSS 348.  {\scshape allesfitter} with Nested Sampling is used to compare different models.  We find a single global likelihood maximum for the best-fit four spot model.  Left panel: the phase-folded photometric data (blue error bars) and 20 posterior models (red curves).  Right panel: the best-fit spot configuration for one posterior sample.}
\label{fig:fit1}  
\end{figure*}

\begin{figure*}
\centering
\includegraphics[width=0.8\textwidth]{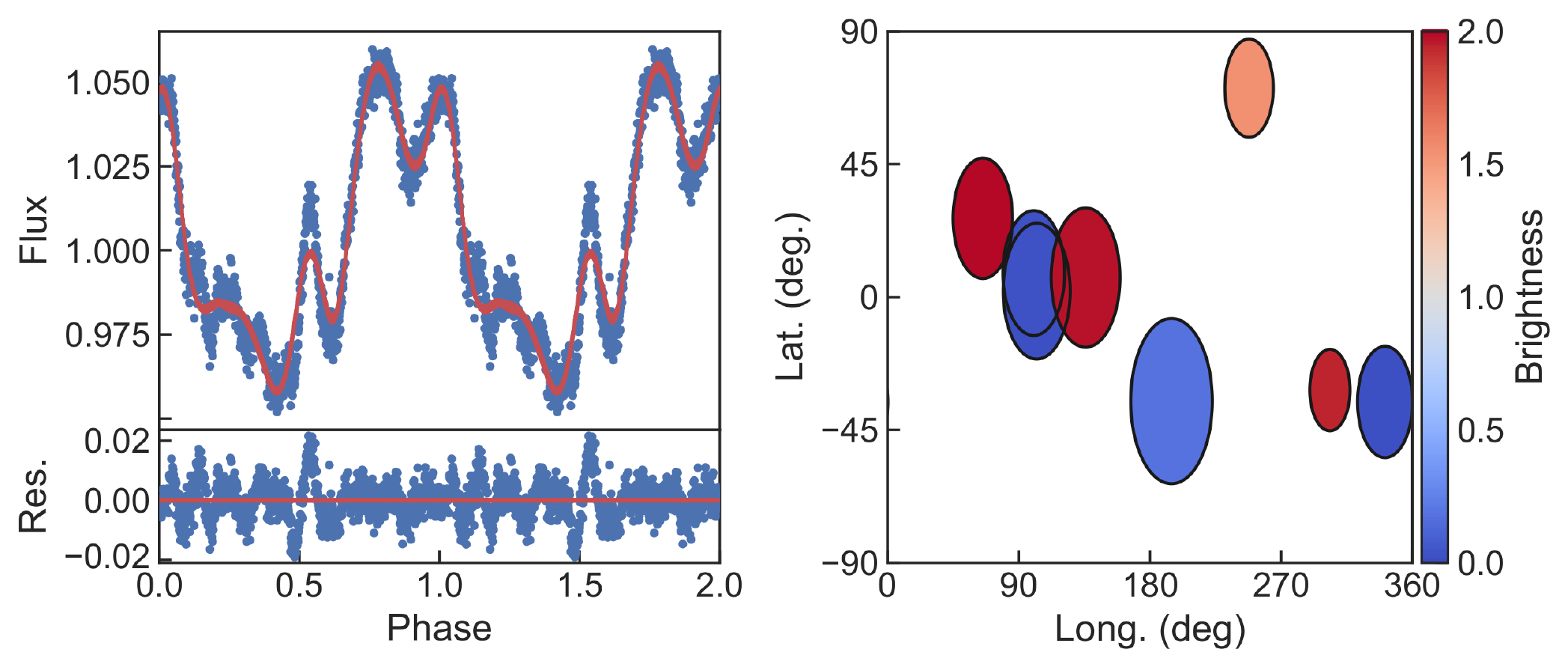}
\caption{A model comprising eight spots cannot adequately describe the periodic variability of TIC 201789285.  While a simpler spot model can account for some overall trends, the sharp, high-frequency features of TIC 201789285 need to be explained by a different process. {\scshape allesfitter} with Nested Sampling was used to compare different spot models.  Left panel: the phase-folded photometric data (blue error bars) and 20 posterior models (red curves).  Right panel: one posterior sample of the best-fit spot configuration.  We note that this spot configuration is locally non-physical, as overlying spots lead to a local stellar surface area that emits a negative flux. This stems from the model trying to replicate the sharp, non-spot-like features.}
\label{fig:fit2} 
\end{figure*}

{\scshape allesfitter}\footnote{\url{https://github.com/MNGuenther/allesfitter}, online Jan 14, 2019} is a publicly available, user-friendly software package built on the packages {\scshape ellc} (light curve and radial velocity models; \citealt{maxted16}); {\scshape aflare} (stellar flares; \citealt{davenport14}); {\scshape dynesty} (static and dynamic Nested Sampling\footnote{\url{https://github.com/joshspeagle/dynesty}, online Jan 14, 2019}); {\scshape emcee} (Markov Chain Monte Carlo sampling; \citealt{foremanmackey13}); and {\scshape celerite} (GP models; \citealt{foremanmackey17}).

{\ron We note two intrinsic approximations in the following fits using {\scshape allesfitter}.  First, broad-band photometric observations, such as those represented by \textit{TESS} lightcurves, do not allow us to discern any wavelength-dependency of individual stellar surface features (e.g., warm plages vs.~cool spots). Second, our approach describes a mean model, averaging over the lifetime and variability time-scales of individual features (which can range from days to weeks).}

\subsection{Verification of our model on HSS 348}
We verify our ability to fit spot models with {\scshape allesfitter} using the example of HSS 348, which shows one of the more complex spot configurations modeled in the literature \citep{strassmeier17}. HSS 348 is a B8 + B8.5 double-lined spectroscopic binary system.  For comparison, we adopt the authors' assumption that all spots are on one star.  We mask out the eclipse region, and phase-fold the {\it CoRot} data on the period reported in \citet{strassmeier17}.

We ran two models, each containing four dark spots, on the phase-folded data.  In the first model, the spin axis of the spotted star is fixed,  according to the inclination of the binary system which is known from eclipse observations, to be normal to the binary orbital plane.  The dilution is fixed to $50\%$ to account for the light of the other star.  We fit for 18 free parameters and adopt the following priors: four parameters for each of the four dark spots (longitude $\mathcal{U}(0\arcdeg,360\arcdeg)$\footnote{A uniform probability distribution from $a$ to $b$ is denoted by $\mathcal{U}(a,b)$.}, latitude $\mathcal{U}(-90\arcdeg,90\arcdeg)$, size $\mathcal{U}(0\arcdeg,30\arcdeg)$, relative brightness $\mathcal{U}(0,1)$), a linear limb darkening term $\mathcal{U}(0,1)$, and the logarithm of a parameter for scaling the photometric error bars $\mathcal{N}(-8,1)$ truncated to $(-23,0)$\footnote{i.e. a normal distribution with mean of -8 and standard deviation of 1, which is truncated to the range from -23 to 0.}.  A small baseline offset taken to be the mean of the residuals is subtracted as part of each step in the sampling.

In the second model we keep the previous settings except that we relax the assumption of a fixed spin axis and fit for the additional parameter, the cosine (prior $\mathcal{U}(0,1)$) of the angle between the axis of the spotted star and the plane of the sky.

Both spot models provide an excellent fit to the variability patterns of HSS 348. We compute the Bayesian evidence for each model in order to compare them quantitatively. We find significant Bayesian evidence for the model with the free spin axis. The best fit of the phase-folded profile and spot configuration for this model are shown in Fig.~\ref{fig:fit1}. 

A casual comparison of our distribution of spots on the stellar surface with that of \citet{strassmeier17} indicates some differences. This is not surprising, as those authors state that their spot solution is not mathematically unique. However, the difference lies basically in the fact that our modeling allows a different temperature for each spot while \citet{strassmeier17} assigned a temperature 1000~K below that of the stellar effective temperature to all four spots. A single spot in each quadrant of stellar longitude is sufficient to reproduce the light curve fairly well in both solutions, including the four observed minima.  An advantage of using Nested Sampling is that it allows one to search for multiple local and global maxima in the likelihood function. The solution shown in Fig.~\ref{fig:fit1} is in fact the only global likelihood maximum for our model.

\subsection{Spots alone cannot describe the rapid rotators}

Using models of 2, 5 or 8 spots, which can here be either bright or dark, we try to describe the complex periodic variations of the rapid rotators using the {\scshape allesfitter} code.  We apply this to the light curve data extracted by the SPOC pipeline.  We do not apply any detrending procedure other than correcting for the dilution of known background stars.
Each light curve is phase-folded on the prominent period identified for the respective object.  We fit for the following parameters with priors: four parameters for each of the 2, 5, or 8 spots (longitude $\mathcal{U}(0\arcdeg,360\arcdeg)$, latitude $\mathcal{U}(-90\arcdeg,90\arcdeg)$, size $\mathcal{U}(0\arcdeg,30\arcdeg)$, relative brightness $\mathcal{U}(0,2)$), the cosine of the spin axis inclination $\mathcal{U}(0,1)$, a linear limb darkening term $\mathcal{U}(0,1)$, and the logarithm of the parameter for scaling the photometric error bars $\mathcal{N}(-8,1)$ truncated to $(-23,0)$.  At each step in the sampling, the mean value of the residuals is computed and subtracted to remove any small baseline offset.

Fig.~\ref{fig:fit2} shows the results of the fit of eight spots to the rotational profile for TIC 201789285.  
The Bayesian evidence favors the 8-spot model over the 2-spot and 5-spot models. This is likely due to the fact that none of the models yields an entirely adequate fit - the eight spots merely give the most freedom to ``bend" the model curve to fit the sharp, high-frequency features.
We note that the spot configuration shown in Fig.~\ref{fig:fit2} is locally non-physical; in the model, overlying spots lead to a local stellar surface area with negative flux. 

In conclusion, even complex spot configurations cannot describe the sharp features found in the phase-folded light curves of these rapid rotators. While spots might contribute to the overall variability, additional physical effects close to the stars must be involved in the formation of the rotational profiles.

\section{Magnetically Constrained Dust}
\label{sec:magnetic}

\citet{stauffer17,stauffer18} have suggested that the highly structured rotational modulation patterns in their sample of Upper Sco and $\rho$ Oph stars are produced by ``orbiting clouds of material at the Keplerian corotation radius''.  Whatever produces the high frequency structure in the phase-folded profiles of those stars is likely to also apply to the stars we report on in this work.  \citet{stauffer17,stauffer18}  go on to say that ``we attribute their flux dips as most probably arising from eclipses of warm coronal gas clouds''.  In this section we re-examine the possibility that absorption by dust rather than scattering by warm gas may be responsible for the dips.

In order for dust to be the cause of the sharp rotational modulations, three requirements must be met: (1) there must be a sufficient amount of dust to account for the observed attenuation in flux; (2) the dust must be able to survive for at least days to weeks before sublimating; and (3) the dust must located where it will corotate with the star, perhaps by virtue of being electrically charged and effectively locked onto magnetic field lines of the star (see e.g., \citealt{farihi17}). 

Given that these stars are typically 40 Myr old, it is not obvious how much dust might remain close to each of them.  A dust sheet comprising $\sim$$10^{17}$ g in submicron dust particles is required to block $\sim$5-10\% of a star's light.

Regarding the third point, the gyroradii of the charged dust particles must be much smaller than orbital distance of the grains, and no larger than the stellar radius. The radius of gyration is:
\begin{equation}
R_g = \frac{mv_\perp c}{qB}  ~({\rm cgs})
\end{equation}
where $m$ and $q$ are the mass and charge of the dust grain, $v_\perp$ is the velocity perpendicular to the magnetic field of strength $B$, and $c$ is the speed of light. If expressed as a ratio with respect to the stellar radius, the requirement becomes
\begin{eqnarray}
\frac{R_g}{R_*} & \simeq & \frac{4 \pi c \rho a^3}{3 Ne} \xi \sqrt{\frac{2GM_*}{d^3}} \frac{d^4}{B_s R_*^4} \ll 1  \label{eqn:gyro} \\
\frac{R_g}{R_*} & \simeq & 0.06\, \xi_{0.1} \, a_{\mu {\rm m}}^3 \, N_{1000}^{-1} \,B_{\rm 3kG}^{-1} P_{\rm crit,hr}^{-1} \left(\frac{d}{R_*}\right)^{5/2} \ll 1 \nonumber 
\end{eqnarray}  
where, in the top equation, $a$ is the grain radius, $\rho$ the bulk density of the grains, $N$ the number of electron charges on each grain, $e$ the electron charge, $M_*$ the mass of the host star, $\xi$ is $v_\perp$ in units of the free-fall speed, $P_{\rm crit}$ the rotational breakup period of the host star, $B_s$ the surface magnetic field (assumed dipolar), and $d$ the distance of the grain from the host star.  In the bottom equation the expression has been evaluated for plausible values, e.g., the grain size in microns, number of charges per grain in units of 1000, $B_s$ the surface field in units of 3 kG, $P_{\rm crit}$ in units of hours, $\xi$ normalized to 0.1, and $\rho$ is taken to be 3 g/cc.  For the somewhat arbitrary normalizations used for these parameters, and for $d/R_* \sim 1$ (i.e., near the stellar surface), it is likely that the dust grains can be shepherded.

Regarding magnetic field strengths, the rapidly rotating and flaring M dwarf star V374 Peg is similar to the stars investigated in the present paper. Its rotational period is 0.4457 days = 10.70 hours, while its mass and radius are 0.30~$M_\odot$ and 0.34~$R_\odot$, all within the parameter ranges of the stars in Table~\ref{tbl:weird}. V374 Peg has a low amplitude stable light variation with some minor changes over time, and has exhibited many flares as well as a possible coronal mass ejection (see \citealt{vida16} for details). More importantly, this star may have a stable global poloidal magnetic field as modeled by \citet{morin08}. The magnetic field of V374 Peg, in locations, reaches strengths of 3-4 kG, implying an overall 1.5-2 kG field of the star (see \citealt{donati06}). New measurements by \citet{shulyak17} reveal a 5.3$\pm$1~kG field on the star.  {\ron Interestingly, the T-Tauri star MN Lupi appears to have nearly the same rotation period, a large magnetic field of several kG, and an active accretion disk connected to the star via magnetic field lines \citep{strassmeier05}.}

\begin{figure}
\centering
\includegraphics[width=1.0\columnwidth]{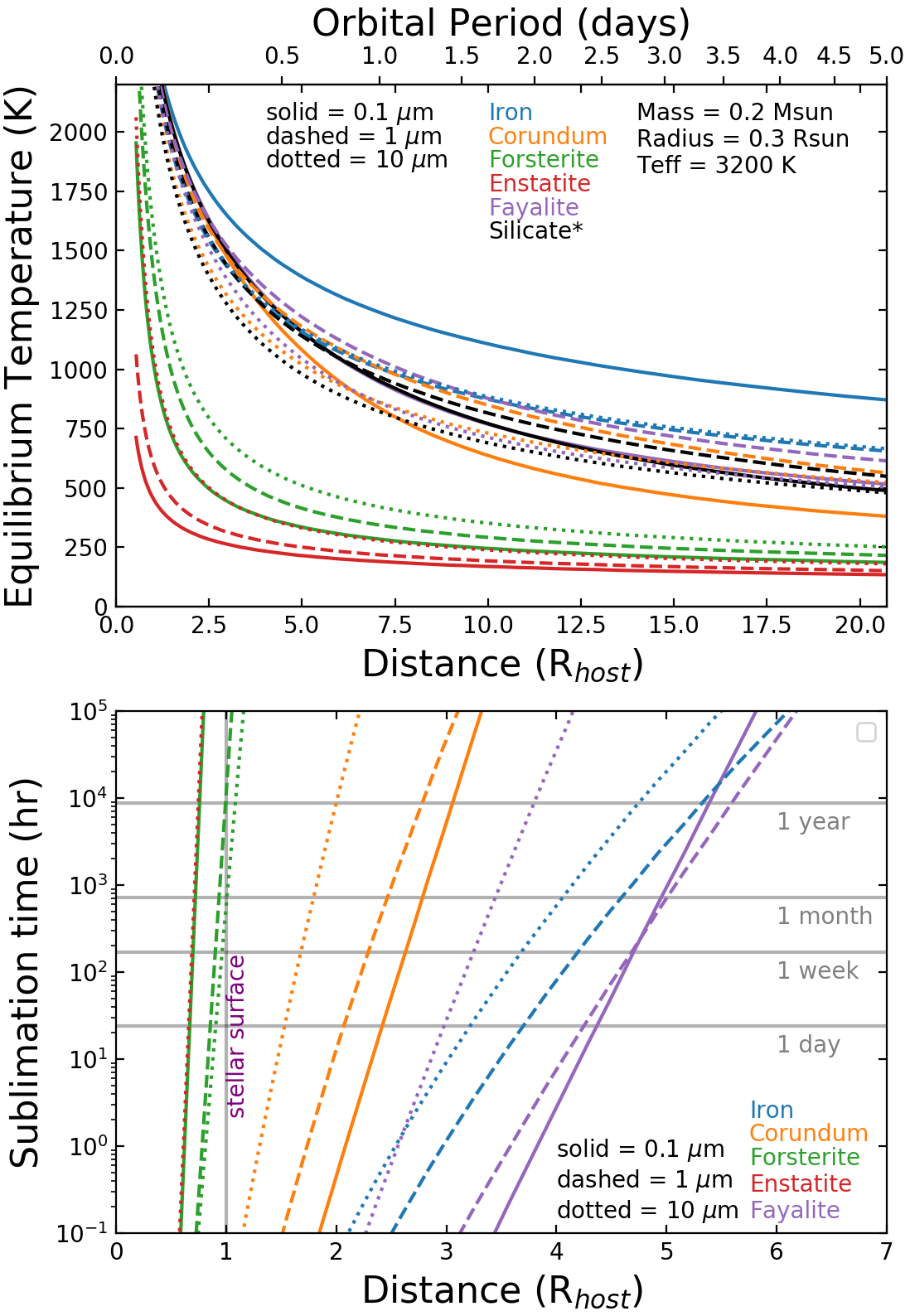}
\caption{Dust grain properties as a function of their distance in units of host star radius from a ``standard'' M star (see text).  The corresponding Keplerian orbital periods are also indicated. Top panel: equilibrium temperature; bottom panel: sublimation lifetime. The color coding legend describes the five minerals being considered; black on the top panel only indicates the ``astronomical silicate" mixture of \citet{draine85}. Continuous, short-dashed, and long-dashed curves are for particle sizes of 0.1, 1, and 10 $\mu$m, respectively.  Dust grains comprised of Mg-rich silicates (e.g., forsterite and enstatite) can survive for substantial periods quite close to the host star.}
\label{fig:grains} 
\end{figure}  

We next assess whether dust grains could survive for substantial times, e.g., a few days, close enough to the M stars that we are considering in this work. We do this in two steps by (1) computing the equilibrium dust-grain temperatures, $T_{\rm eq}$, for unshielded grains in circular orbits at each of a range of distances, i.e., orbital radii, from 1 $R_*$  to 30 $R_*$, and (2) using $T_{\rm eq}$ and the bulk properties of the grain material to compute the sublimation time, $\tau_{\rm sub}$. We do this for grains of three different sizes (0.1, 1, and 10 $\mu$m) and 6 different compositions revolving around a canonical M star with $M_* = 0.3 \, M_\odot$, $R_* = 0.4 \, R_\odot$, and $T_{\rm eff} =3200$ K.  In order to compute $T_{\rm eq}$ we solve the thermal balance equation by matching the emission of radiation at $T_{\rm eq}$ to the absorption of radiation from the host star at $T_{\rm eff}$ (we ignore the cooling effects of sublimation as this effect is rarely important).  To solve this equation we utilize the grain absorption cross sections based on Mie scattering theory following \citet{croll14} and \citet{xu18}.  In turn, the Mie calculations require, as input, the real and imaginary parts of the indices of refraction ($n$ and $k$, respectively) of the dust grains (taken to be spherically shaped).  We study 6 minerals that have been considered for the tails of the so-called ``disintegrating exoplanets'' (see, e.g., \citealt{vanlieshout18} for a review).  These include iron \citep{querry85}, corundum \citep{begemann97}, forsterite \citep{jager03}, enstatite \citep{jager03}, and fayalite\footnote{\url{https://www.astro.uni-jena.de/Laboratory/Database/jpdoc/f-dbase.html}}, where the citations in parentheses are original sources of the corresponding $n$ and $k$ values.  The sixth mineral is a generic ``astronomical silicate" mixture \citep{draine85}. We obtained the indices of refraction from the HITRAN aerosol database \citep{2017JQSRT.203....3G} and bulk mineral properties, such as the vapor pressure vs.~$T_{\rm eq}$, from Table 5 of \citet{vanlieshout16}.
		
The results of this grain survival study are shown in Fig.~\ref{fig:grains}.  The top panel shows the equilibrium grain temperatures as a function of distance from the host star out to 20 $R_*$; the corresponding Keplerian orbital periods are labeled along the top axis.  For distances of $\lesssim 5 \,R_*$, the temperatures of the grains of iron, corundum, and fayalite (as well as cosmic silicates) are modestly high, while for forsterite and enstatite (the Mg-based silicates) the temperatures can be surprisingly low.  This is due to the fact that these latter two minerals have low values of $k$ ($\sim$ 0.001) near the wavelengths they are absorbing and relatively high values of $k$ ($\sim$ 1) at the much longer emission wavelengths.  Grain sublimation times for 15 combinations of grain sizes and compositions that we consider are shown as a function of distance in the bottom panel of Fig.~\ref{fig:grains}. (Sublimation times for the astronomical silicate mixture are not given since there is no single set of mineral properties for this collection of materials.)  We see that grains with sizes in a relevant range and composed of a number of kinds of minerals can survive for more than a day at distances of $\lesssim 5 R_*$, while grains of 3 of 5 mineral types can survive at distances of $\lesssim 3 R_*$; 2 of 5 can even survive relatively close to the stellar surface. 

Perhaps the biggest difficulty with the dust-grain model, or any other model invoking obscuration by intervening material (including coronal gas), lies in the distance from the host star where the material must be located. In order for any attenuating material to produce narrow duty-cycle features in the rotational light curve it must be situated considerably above the stellar surface.  If $f$ is the full width of the feature divided by the rotation period, the distance, $d$, from the stellar center, in units of the radius of the host star, $R_*$, must be
\begin{equation}
d \gtrsim R_* \csc(\pi f) .
\label{eqn:dutycycle}
\end{equation}
If, for example, we consider narrow features with duty cycles, $f$, of 1-10\% (see, e.g., Figs.~\ref{fig:rawLC}, \ref{fig:folds}, and \ref{fig:changes}), then $d$ should be $\gtrsim 3-30\,R_*$.  (This is basically the reason that starspots, with $d/R_* =1$ do not work.)  Such large factors of $d/R_*$ in Eqn.~(\ref{eqn:gyro}), raised to the 5/2 power, increase the leading coefficient from 0.06 to values of $\sim$1 to 300 which quickly violates the requirement for meaningful constraint by the magnetic field.

\section{A Dusty Ring Model}
\label{sec:ring}

We now consider a model where the star is orbited by one or more rings composed of dust-size or somewhat larger particles.  In this model, there would be no requirement for dust to be magnetically entrained in the stellar magnetic field at distances of many stellar radii where the magnetic field would be relatively weak as in the magnetically constrained dust model discussed in Sect.~\ref{sec:magnetic}.  The ring particles would move in Keplerian orbits at relatively large distances from the star, and therefore the sublimation lifetime would not be an issue even if the particles are dust-like in size.  Furthermore, there would be no requirement for small coherent dust structures to form and be maintained.  

\begin{figure}
\centering
\includegraphics[width=1.0\columnwidth]{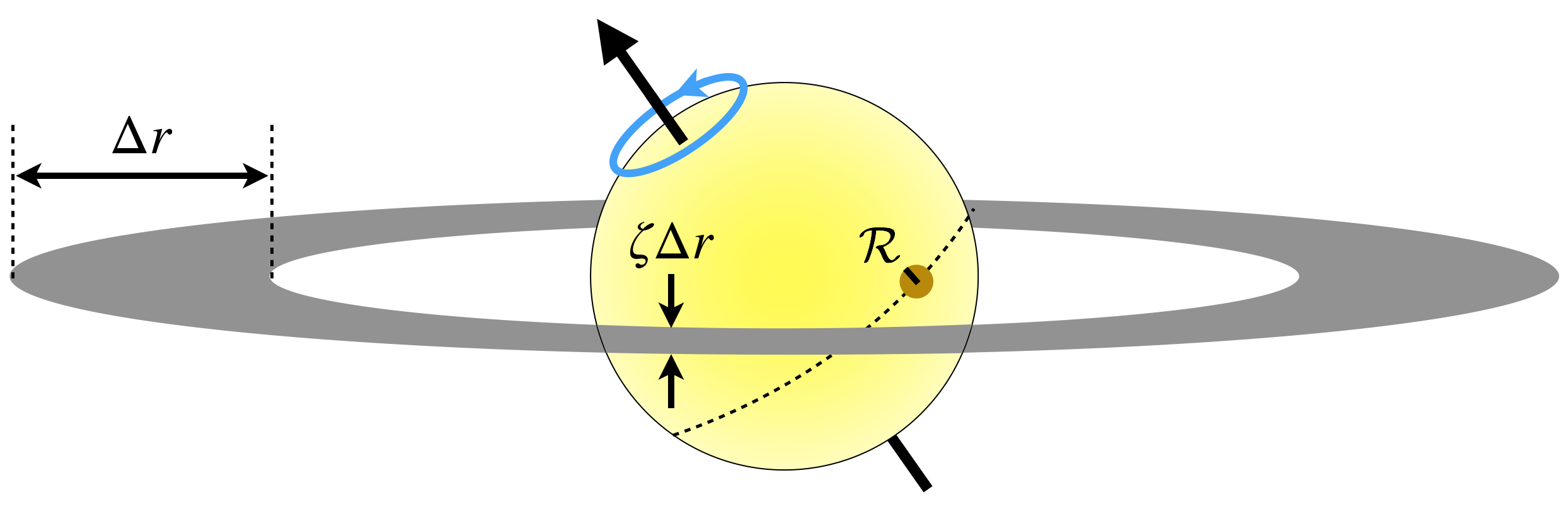}
\caption{Sketch of a dusty ring geometry that might produce sharp modulation features. The disk plane is tilted by a small angle $\zeta$  with respect to the edge-on orientation, and the stellar spin axis is oriented such that the stellar rotation will carry spots behind the disk for short time intervals. The radius of a stellar spot is denoted as $\mathcal{R}$.}
\label{fig:ringsketch} 
\end{figure}

A sketch of the envisioned geometry is shown in Fig.~\ref{fig:ringsketch}.  In terms of radial structure, the disk could be either mostly continuous with some gaps or thin regions, or it could comprise a small number of concentric rings.  The ring (or rings) would be viewed at a tilt angle with respect to the edge-on orientation, $\zeta$, that is small enough for the ring to partially occult the star.  For a range of orientations of the star's rotation axis such as that shown in the figure, cool or warm spots will be hidden by the tilted rings for short time intervals as the star rotates and thereby produce rapid photometric changes.

Consider first just a single ring extending from radius $r$ to $r+\Delta r$, or a gap in a more continuous ring of the same dimensions.  We assume that the disk is thin, i.e., having a half-thickness $h$ that satisfies $h \ll \Delta r$.  With this configuration, there would be a nearly linear feature across the star of extra transmission or attenuation having a thickness $\zeta \Delta r$ and a vertical displacement from the center of the star of $\zeta r$.  A requirement for this gap to produce a sharp temporal feature as the star and its spots rotate beneath it, is that $\zeta \Delta r \ll R_*$.  A final requirement is the existence of stellar surface features that can produce photometric modulations of order 1\% by this mechanism.  In turn, this implies the existence of regions with characteristic sizes $\mathcal{R}$ where $\mathcal{R} \sim \zeta \Delta r \sim 0.1R_*$, and with surface brightnesses that are sufficiently different from the mean stellar surface brightness to produce $\sim$1\% effects.

Finally, in regard to a possible occulting ring, we can estimate a lower bound on its mass.  The mass of a single uniform ring is given by:
\begin{equation}
\mathcal{M}_{\rm ring} \simeq \pi \left(R_{\rm out}^2-R_{\rm in}^2\right) \Sigma
\label{eqn:diskmass}
\end{equation}
where $R_{\rm out}$ and $R_{\rm in}$ are the outer and inner radii of the ring, and $\Sigma$ is the mass column density of the particles comprising the ring in the direction normal to the ring plane.  An illustrative set of parameters that could produce a sharp feature of width a few percent of the modulation period with $\sim$1\% amplitude would be: $R_{\rm in} = 10\,R_*$, $R_{\rm out} = 15\,R_*$, $\mathcal{R} = 0.2\,R_*$, and $\zeta =0.04 \simeq 2^\circ$.

A ring consisting of micron-size dust particles with a projected number column density $\sim$$1 \, \mu {\rm m}^{-2}$ would involve a minimum total mass to provide the desired light attenuation. In that case, for grains whose bulk material density is $\sim$3 gm/cc, the mass of a grain is $\sim$$10^{-11}$ g, and $\Sigma \approx 0.001 \,\zeta \, {\rm g~cm}^{-2}$. Plugging this value for $\Sigma$, the above illustrative values of $R_{\rm in}$, $R_{\rm out}$, and $\zeta$, and $R_* \sim 0.3\, R_\odot$ into Eqn.~(\ref{eqn:diskmass}), we find a mass of $\approx 10^{19}$ grams. The mass of the disk ring would be larger for larger particle sizes, $\Delta r$, or $\zeta$. 

{\ron Finally, we estimate the precession time for an inclined disk with the above parameters.  We take the apsidal motion constant, $k_2$, for the M stars to be that of an $n=3/2$ polytrope with a value of $k_2 \simeq 0.14$ \citep{brooker55}.  From that, we compute the $J_2$ quadrupole term to be in the range of 0.001 to 0.024 for stellar rotation periods in the range of 1.6 to 4 h.  The resultant disk precession times are in the range of 10-70 yr, even for the shortest periods among the highly structured rotators.}

\begin{figure*}
\begin{center}
\includegraphics[width=0.7\textwidth]{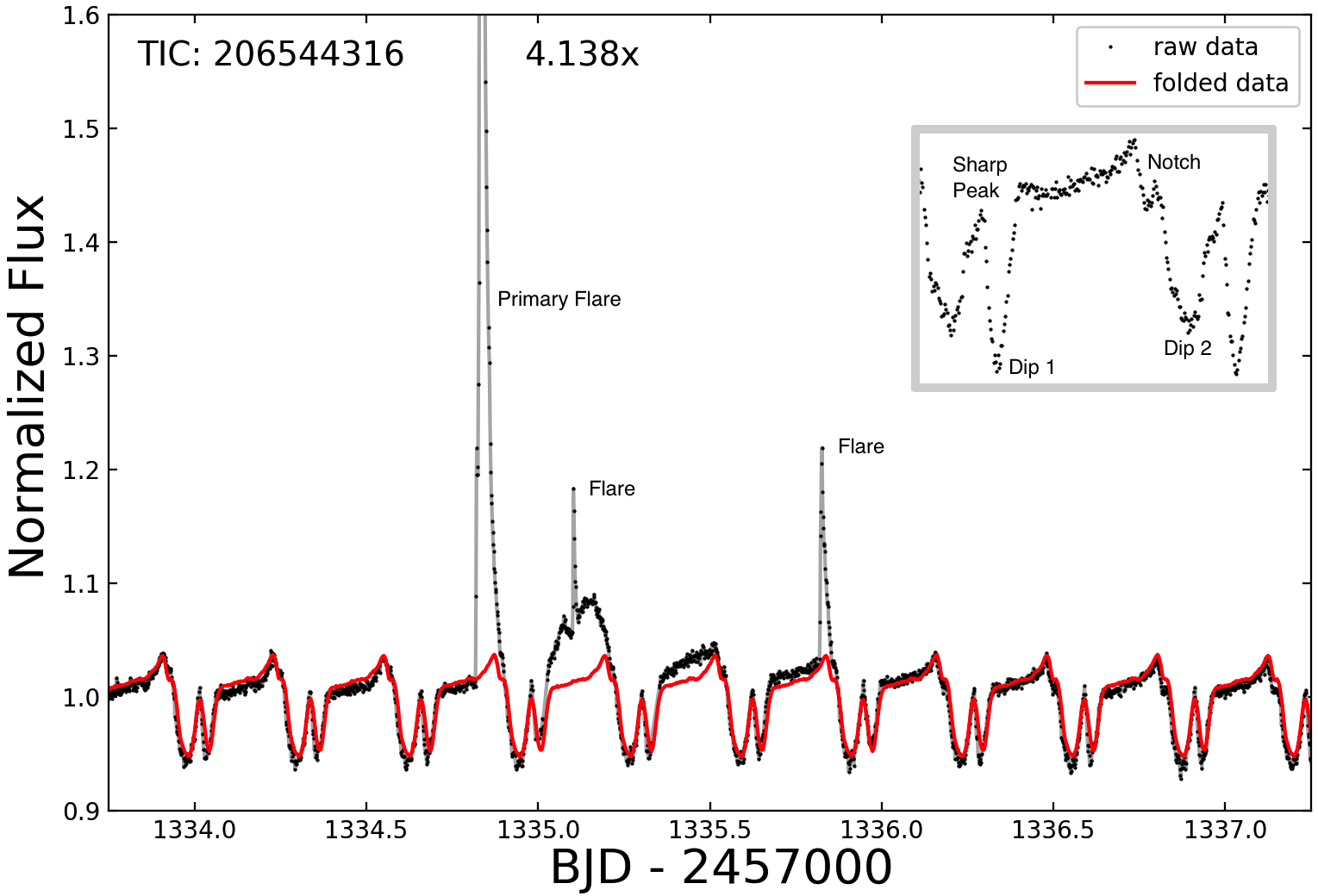}
\caption{The light curve of TIC 206544316 showing changes to the rotational modulation pattern following a large flare and around the times of two subsequent smaller flares.  The black points are the normalized fluxes with 2-minute cadence.  The over-plotted red curve is the average rotational modulation pattern for this source.  Substantial deviations from the mean profile are seen for three rotational cycles following the large flare. The inset panel shows a zoom in on one of the modulation cycles that occurs well after the large flare and is annotated to define various features discussed in the text. }
\label{fig:flare}
\end{center}
\end{figure*} 
 
\section{Radiation Beaming}
\label{sec:beaming}

Beaming of the outgoing radiation is possibly another way of producing sharp features in rotational profiles.  This could come about if radiation passing through magnetized plasma near the stellar surface has an anisotropic transmission factor.  

\citet{hamada74} have computed how the Thomson scattering cross section is modified in the presence of a strong magnetic field.  For an unpolarized beam, incident at angle $\theta$ with respect to the magnetic field direction, the total scattering cross section is \citep{hamada74}:
\begin{equation}
\sigma = \sigma_{\rm Th} (1-u)^{-2} \left[1+u+(u/2)(u-3)\sin^2 \theta \right]
\label{eqn:sigma}
\end{equation}
where $u$ is the ratio $\nu_C/\nu_{\rm rad}$, $\nu_C$ is the cyclotron frequency, $\nu_{\rm rad}$ is the frequency of the radiation, and $\sigma_{\rm Th}$ is the classical Thomson cross section.  The cyclotron frequency is: $\nu_C \simeq 3 \,B_{\rm kG} \, {\rm GHz}$ while the frequency of the radiation in the {\em TESS} band is $\approx 3 \times 10^{14}$ Hz.  Since $u$ is obviously quite small, we can expand Eqn.~(\ref{eqn:sigma}) in a Taylor series to show that:
\begin{equation}
\sigma \simeq \sigma_{\rm Th} \left[ 1 +3 \left(1-\sin^2 \theta /2\right)\,u \right]
\label{eqn:sigma2}
\end{equation}
Therefore, the magnitude of the difference in the cross section between radiation incident in the direction of the magnetic field vs.~perpendicular to it is of the order $u \sim 10$ ppm, and therefore appears to be unable to lead to significant beaming.  

Barring some other type of physical effect that can yield beaming that can, in turn, produce a $\sim$5\% amplitude with beaming angles smaller than $\sim$20$^\circ$, we do not see how beaming can produce the effects we see.

\section{Large Flares and Changes to the Rotational Modulation Pattern}
\label{sec:flare}

The large flares that are sometimes observed in these stars may provide clues to the origin of the complex rotational modulation patterns. In fact, \citet{stauffer17, stauffer18} found that the rotational modulation profile changed significantly after flares were observed in some of the systems in Upper Sco.  This prompted us to examine the behavior of our 10 rotators with structured modulation profiles after the occurrences of flares.  In the {\em TESS} light curves, large flares where the intensity was seen to increase by more than a factor of two were observed on only two of our 10 stars, and, of these, a distinct change in modulation profile was seen only in the light curve of TIC 206544316 after a large 4$\times$ superflare \citep[$E_\mathrm{bol}=1.7 \times 10^{34}$~erg;][]{gunther19}.

The light curve of this star around the time of this event is shown in Fig.~\ref{fig:flare}.  In this figure the average 0.32-day complex modulation pattern is shown in red, while the photometric measurements are shown in black.  After the 4$\times$ flare, the three subsequent rotational profiles have larger amplitudes and somewhat different shapes.  Two additional smaller flares erupt later, one during the next profile that has a strongly enhanced maximum brightness, and another three cycles later, again at higher maximum light than before or well after the big flare.

About 25 rotations later than the flares shown in Fig.~\ref{fig:flare} another series of three or four flares with smaller peak brightnesses are observed within an interval of 5 rotations, and they occur at rotational phases similar to those seen in Figure~\ref{fig:flare}.

The two smaller flares seen in Fig.~\ref{fig:flare} are likely `sympathetic flares', that is, flares that sometimes appear, not merely by coincidence in time, soon after a large flare in analogy to such events that have been observed on the Sun (for the solar case see \citet{torok11}). Such temporally-correlated solar flares may be produced subsequent to disturbances of the large-scale magnetic field of the Sun from certain flares, as particular types of disturbances of the field may give rise later to other events at other locations on the Sun (see \citet{schrijver11}). TIC~206544316 has parameters similar to those of V374~Peg that is inferred to have a strong and stable global magnetic field (see Sect.~\ref{sec:magnetic}). Thus we can suppose there is a similar strong, global magnetic field on TIC~206544316 which, after being disturbed, could induce additional flaring events. 

In light of the above, we interpret the disturbance to the light curve of TIC 206544316 in Fig.~\ref{fig:flare} in the following way. First, we have fit the large flare by the same model discussed in \citet{gunther19} and find it to be consistent with the flare having decayed away by the time of the subsequent rotation of the host star. Second, the two prominent dips in the light curve are mostly unaffected by the flare for the next few rotations. 
However, another smaller feature (the `notch'), just after the maximum in the rotational profile, changes permanently after the flare.
Third, we believe that the excess amplitude of the profiles in the three rotations following the flare may be attributed to the aftermath of the flare.

Aside from this one dramatic event in TIC 206544316 we find no other case where a flare appears to be associated with a distinct change in either the shape or amplitude of the highly structured rotational profile of any of the other 9 stars of our set.

In the dust model, a large flare could at least temporarily change the modulation profile of the star by, e.g., inducing changes in the configuration of the magnetic field, by increasing the rates at which grains sublimate, or by inducing changes in the amounts of net electrical charge on the grains.  Likewise, in the model noted by \citet{stauffer17}, which invokes corotating coronal gas as a scattering mechanism, a large flare may also be expected to result in temporary changes in the modulation profiles.  Our observation of changes in the modulation profile of one star following one flare are not sufficient to give us insight into the physical mechanisms underlying the modulations, and certainly on its own does not confirm or contradict any of the models we have considered. 

\section{The Stellar Associations}
\label{sec:associations}

Rapidly rotating M dwarfs are often young and, consequently, close to the stellar associations into which they were born.  In such cases, the stellar ages may be estimated from the ages of the stellar associations.  To try to figure out which stellar association each of our target stars might belong to, we used the online BANYAN $\Sigma$ Bayesian analysis tool relevant to nearby young associations \citep{gagne18}.  If provided sufficient information about the 6-dimensional phase-space coordinates of a star (Table \ref{tbl:weird}), BANYAN $\Sigma$ assesses the likelihood that the star belongs to one of the 27 known and well-characterized young associations within 150 pc of the Sun. 

Each of our 10 objects is most likely to be a member of the stellar association given in Table \ref{tbl:weird}.  Five out of 10 belong to the association Tucana Horologium.  For convenience, we give in Table \ref{tbl:associations} the names, distances, and ages of the five associations listed in Table \ref{tbl:weird}. These associations all have distances between 30 and 50 pc and the ages of four of the five are between 25 and 50 Myr. 

\begin{table}[]
\begin{tabular}{llcc}
\hline
Name 			& Abbr. 		& Ave. Dist. [pc]	& Age [Myr] \\
\hline
AB Doradus 		& ABDMG 	& $30^{+20}_{-10}$ 	& $149^{+51}_{-19}$ \\
$\beta$ Pictoris 	& $\beta$PMG 	& $30^{+20}_{-10}$	& $24\pm3$ \\
Carina 			& CAR 		& $60\pm20$ 		& $45^{+11}_{-7}$ \\
Columba 			& COL 		& $50\pm20$ 		& $42^{+6}_{-4}$ \\
Tucana Horologium 	& THA 		& $46^{+8}_{-6}$ 	& $45\pm4$ \\
\hline
\end{tabular}
\caption{Data from \citet{gagne18} }
\label{tbl:associations}
\end{table}

These ages of $\sim$40 Myr are considerably older than the nominal 10 Myr age of the Upper Scorpius association \citep{pecaut12,feiden16} which houses many of the known dippers as well as the structured rotators found by \citet{stauffer17,stauffer18}. It is not clear whether it is reasonable or not to postulate that 40-Myr-old M stars can accrete sufficient amounts of dust into their magnetospheres to produce the observed rotational modulations.

\section{Summary and Discussion}
\label{sec:summary}

In this work we have reported the discovery, using the first two sectors of {\em TESS} data, of 10 new rapidly rotating M-dwarfs with highly structured rotational modulation patterns.  The overall shapes of the periodic patterns can remain stable for weeks, while individual features can change within days.  As part of this study, we also report the discovery of {\NSecTotalFastMdwarf} rapidly rotating M-dwarfs with periods $< 4$ hr, of which the shortest period is 1.63 hr. These objects are part of a sample of 371 rapidly rotating M-dwarfs also found in the first two sectors of {\em TESS} data with $T_{\rm eff}$ cooler than 4000 K and with rotation periods $<$ 1 day.

We attempt to better understand the nature of the rapidly rotating M-dwarfs with highly structured rotational modulation patterns, which are similar in their properties to a number of stars found in the Upper Sco stellar association \citep{stauffer17}. First, we fitted general star-spot models, able to accommodate large numbers of spots, to these periodic variations which exhibit large amounts of structure (Sect.~\ref{sec:spots}).  However, we show that spots are not likely to be able to explain the sharp features in the light curves (see also \citealt{stauffer17}). Next we explored the possibility that the highly structured rotational modulation patterns are due to absorption by charged dust particles that are trapped by the stellar magnetosphere (see Sect.~\ref{sec:magnetic}). We have shown that dust grains may be able to survive for a number of rotation periods in the magnetosphere at the corotation radius.  However, such dust would have to be substantially far from the host star (see Eqn.~\ref{eqn:dutycycle}) to produce the sharp features observed in the rotation profiles, and the existing magnetic fields at those distances would likely be too weak to trap charged dust particles (see Sect.~\ref{sec:magnetic}).  Beaming of the emitted radiation would be a desirable mechanism, but we show, at least in the case of electron scattering in a magnetic field, that the observed amplitudes of a few percent are not achievable (Sect.~\ref{sec:beaming}).  We did not explore the scenario, proposed by \citet{stauffer17}, involving attenuation by scattering from warm coronal material, because it is difficult to conceive of a situation where it would produce significant emission or absorption.  

Perhaps the most natural of the scenarios we have explored is that of a spotted host star rotating obliquely inside of a dusty ring of material a few tens of stellar radii away (Sect.~\ref{sec:ring}).  While potentially very promising, this model has not been fully fleshed out, and a determination of whether it can quantitatively explain the sharp modulation structures that are observed will require further exploration.  However, this is beyond the scope of the present work. 

In addition, we explore whether large flares could serve as probes of the highly structured rotational modulation patterns. We found 32 large flaring events with flux increases larger than 2 times the stellar flux. However, we do not see any changes in the intrinsic rotational modulation pattern after most flares. The exception is a large $4\times$ superflare event ($E_{\rm bol}=1.7 \times 10^{34}$~erg) on TIC~206544316, which is the strongest flare observed to date in such a rapidly-rotating M dwarf with highly structured rotational modulation patterns. This superflare apparently changes the subsequent periodic modulation by increasing the maximum brightness for a few rotations, and altering one feature permanently.  

Because most of our 10 highly structured rotators are in well-studied stellar associations, their ages are known reasonably well.  Therefore, we are able to compare some of the properties of our stars with the `scalloped shell' rotators in Upper Sco and $\rho$ Oph.  The latter have ages of 5 - 10 Myr, distances of about 135 pc, and a median $K_s$ magnitude of 10.5 (median $M_K \simeq 4.9$).  The comparable rotators in this work are some three times closer, and several times older.  Both sets of stars have a similar median $K_s$ magnitude, but a median $M_K \simeq 6.4$, which is perhaps not surprisingly 1.5 magnitudes less luminous than the `scalloped shell' modulators in upper Sco and $\rho$ Oph which are younger.  

A small sample of low mass stars revealing periodic optical variability has also been found to be radio active \citep[see e.g.][]{harding2013} -- however, we note that their optical modulations seem not as sharp-featured as those of our rapid rotators.  For these radio active low mass stars, electron cyclotron maser (ECM) emission is suggested to cause periodic, circularly polarized radio pulses \citep[see e.g.][]{hallinan2007}.  Notably, the ultracool dwarf TVLM 512-46546 shows sharp radio bursts every 1.96 hours \citep{hallinan2007} and smooth optical modulations with a similar $\sim$2~h period \citep{littlefair2008, harding2013}.  The latter was attributed to inhomogeneous dust clouds in the relatively cool stellar atmosphere \citep{littlefair2008}.  This short period is remarkably similar to the periods of our 10 complex rotators, and suggests that the optical and radio emission are linked by common magnetic phenomena.  The modulation signal of TVLM 512-46546, for example, is stable over a time span of five years. Neither spots nor magnetically constrained dust could explain such long-term stability.  Whether the modulations of our targets are as long-lived may be studied in the future.

The mechanism that produces the observed highly structured rotational modulation patterns has not yet been confidently identified.  In that regard, we have discussed, in Section~\ref{sec:ring}, a promising idea.  Continued observation of these 10 complex rotators will also provide opportunities to further understand the relevant physical processes. Ideally, this will encompass observations of time variability in the radio band, in both optical and near-infrared bandpasses, and in H$\alpha$, spectroscopy well beyond that presented in Section~\ref{sec:spectra}), and direct imaging.  Finding $\sim$100 more of these objects in the upcoming {\em TESS} sectors, as we anticipate, will hopefully further elucidate their nature.  

This work was carried out as part of a larger search for rapid rotation and high frequency oscillations in the {\em TESS} data.  The methods described herein can also be applied to other high frequency stellar phenomena.

\acknowledgments

{\ron The authors are grateful to the reviewer, Klaus Strassmeier, for a number of important suggestions which improved the manuscript.}
We acknowledge the use of public data collected by the TESS mission, which are from pipelines at the TESS Science Office and at the TESS Science Processing Operations Center and made publicly available from the Mikulski Archive for Space Telescopes (MAST). Funding for the TESS mission is provided by NASA's Science Mission directorate. 
Resources supporting this work were provided by the NASA High-End Computing (HEC) Program through the NASA Advanced Supercomputing (NAS) Division at Ames Research Center for the production of the SPOC data products.
M.\,N.\,G.~acknowledges support from MIT's Kavli Institute as a Torres postdoctoral fellow.
We are grateful to Lidia van Driel-Gesztelyi for input about solar analogs to some of the observed features of the systems studied in this work.

\textit{Software}:
{\scshape python} \citep{rossum95},
{\scshape numpy} \citep{vanderwalt11},
{\scshape scipy} \citep{jones01},
{\scshape matplotlib} \citep{hunter07},
{\scshape tqdm} (doi:10.5281/zenodo.1468033),
{\scshape seaborn} (\url{https://seaborn.pydata.org/index.html}),
{\scshape allesfitter} {\guentherinprep},
{\scshape ellc} \citep{maxted16},
{\scshape aflare} \citep{davenport14},
{\scshape dynesty} (\url{https://github.com/joshspeagle/dynesty}),
{\scshape corner} \citep{foremanmackey16}.










\end{document}